\documentclass[aip,
 jmp,
 amsmath,amssymb,
preprint,%
 reprint,%
]{revtex4-1}

\usepackage[a4paper, total={6in, 9.5in}]{geometry}
\usepackage{graphicx}% Include figure files
\usepackage{dcolumn}% Align table columns on decimal point
\usepackage{bm}% bold math
\usepackage{amsmath}
\let\Omega\varOmega
\usepackage[utf8]{inputenc}
\usepackage[T1]{fontenc}
\usepackage{etoolbox}
\usepackage{xcolor,soul}
\usepackage{subfigure}
\sethlcolor{yellow}

\makeatletter
\def\@email#1#2{%
 \endgroup
 \patchcmd{\titleblock@produce}
  {\frontmatter@RRAPformat}
  {\frontmatter@RRAPformat{\produce@RRAP{*#1\href{mailto:#2}{#2}}}\frontmatter@RRAPformat}
  {}{}
}%
\makeatother
\begin{document}

\preprint{Submitted to Physics of Fluids}

\title[]{Painting Taylor vortices with cellulose nanocrystals: \\ supercritical spectral dynamics \\ \hfill}
% Force line breaks with \\
\author{Reza Ghanbari}
\affiliation{%
Chalmers University of Technology, Department of Industrial and Materials Science, \\ 42 916 Gothenburg, Sweden%\\This line break forced% with \\
}
\affiliation{%
MAX IV Laboratory, Lund University, Lund, Sweden%\\This line break forced% with \\
}%
 \altaffiliation[Present affiliation: ] {NKT Technology Consulting, Västerås, Sweden}%Lines break automatically or can be forced with \\

\author{Sajjad Pashazadeh}%
\affiliation{ 
Chalmers University of Technology, Department of Industrial and Materials Science, \\ 42 916 Gothenburg, Sweden%\\This line break forced with \textbackslash\textbackslash
}%

\author{Kesavan Sekar}
\affiliation{%
Chalmers University of Technology, Department of Industrial and Materials Science, \\ 42 916 Gothenburg, Sweden
}%

\author{Kim Nygård}
\affiliation{%
MAX IV Laboratory, Lund University, 22 484 Lund, Sweden%\\This line break forced% with \\
}%

\author{Ann Terry}
\affiliation{%
MAX IV Laboratory, Lund University, 22 484 Lund, Sweden%\\This line break forced% with \\
}%

\author{Marianne Liebi}
\affiliation{%
Paul Scherrer Institute, PSI, Villigen 5232, Switzerland%\\This line break forced% with \\
}%
\affiliation{%
Empa, Swiss Federal Laboratories for Materials Science and Technology, Center of X-ray Analytics, St.Gallen, Switzerland%\\This line break forced% with \\
}%
\affiliation{%
Chalmers University of Technology, Department of Physics, \\ 42 916 Gothenburg, Sweden%\\This line break forced% with \\
}%

\author{Aleksandar Matic}
\affiliation{%
Chalmers University of Technology, Department of Physics, \\ 42 916 Gothenburg, Sweden%\\This line break forced% with \\
}%

\author{Roland K\'{a}d\'{a}r}
%\email{Roland.Kadar@chalmers.se.}
 \homepage{Corresponding author: roland.kadar@chalmers.se}
\affiliation{%
Chalmers University of Technology, Department of Industrial and Materials Science, \\ 42 916 Gothenburg, Sweden%\\This line break forced% with \\
}%
\affiliation{%
MAX IV Laboratory, Lund University, Lund, Sweden%\\This line break forced% with \\
}%
\affiliation{%
Wallenberg Wood Science Centre (WWSC), Chalmers University of Technology, \\ 42 916 Gothenburg, Sweden%\\This line break forced% with \\
}%

%\tableofcontents

\date{\today}% It is always \today, today,

\begin{abstract}

%\linenumbers

We study the flow stability and spatio-temporal spectral dynamics of cellulose nanocrystal (CNC) suspensions in a custom Taylor-Couette flow cell using the intrinsic shear induced birefringence and liquid crystalline properties of CNC suspensions for flow visualizations for the first time. The analysis is performed at constant ramped speed inputs of the independently rotating cylinders for several cases ranging from only inner or outer rotating cylinders to three counter-rotation cases. All CNC suspensions have measurable elastic and shear thinning, both increasing with CNC concentration. We show that the flow patterns recorded are essentially Newtonian-like, with non-Newtonian effects ranging from a decrease in wavenumbers to altering the critical parameters for the onset of instability modes. Outer cylinder rotation flow cases are stable for all concentrations whereas inner cylinder rotation flow cases transition to axisymmetric and azimuthally periodic secondary flows. However, unstable counter-rotation cases become unstable to asymmetric spiral modes. With increasing CNC concentration a counter-rotation case was found where azimuthally periodic wavy patterns transition to asymmetric spiral modes. In contrast to polymeric solutions of similar low to moderate elasticity and shear thinning, the shear-thinning region of CNC suspensions is expected to lead to the breakdown of the chiral nematic phase, whose elastic constants constitute the dominant structural elasticity mechanism. Thus, we interpret the Taylor-Couette stability of the CNC suspensions as dominated by their shear-thinning character due to the expected loss of elasticity in nonlinear flow conditions.
\end{abstract}

\maketitle

%\begin{quotation}
%The ``lead paragraph'' is encapsulated with the \LaTeX\ 
%\verb+quotation+ environment and is formatted as a single paragraph before the first section heading. 
%(The \verb+quotation+ environment reverts to its usual meaning after the first sectioning command.) 
%Note that numbered references are allowed in the lead paragraph.
%
%The lead paragraph will only be found in an article being prepared for the journal \textit{Chaos}.
%\end{quotation}

\section{Introduction}

Since their discovery a century ago\cite{taylor1923}, Taylor-Couette (TC) flow, i.e. flow between rotating concentric cylinders \cite{fardin2014hydrogen}, continues to be the most prominent benchmark case for flow stability. This is due to its distinctive complex flow fields in the form of uniquely rich supercritical flow patterns beyond the limit of laminar Couette flow. This has triggered a broad spectrum of experimental and theoretical studies on analyzing emerging flow patterns and their stability in both Newtonian and rheologically complex fluids \cite{andereck_1986,muller1989purelyelastic,avgousti1993non, groisman2004elastic, KADAR2012TC,liu2013polymer,greidanus2015turbulent,bengana2019spirals}.
Elucidating the spatio-temporal features of supercritical flow translations is not only important to explain fundamental nature of nonlinear dynamical systems, but is also relevant to predicting the impact of flow translations and instabilities on the transport phenomena present in technological applications such as chemical reactions \cite{bioreaction2004,bioreaction2019,reaction2021}, drilling \cite{drilling2021}, filtration devices \cite{filtration1989,filtration1991}, the shearing process of proteins \cite{protein2006,protein2009,protein2019} and others \cite{greidanus2015turbulent, TC_application_2015, TC_application_2017}. 

The TC flow of Newtonian fluids has been broadly investigated, with a rotating inner cylinder while the outer cylinder is kept stationary, being the most studied configuration. The onset of different flow patterns has been determined both theoretically and experimentally in terms of the Reynolds ($Re$) and / or Taylor ($Ta$) numbers. $Re$ quantifies for the ratio of inertial to viscous forces, with the $Ta$ number being a modified version of $Re$, $Ta \propto Re^2$ to account for centrifugal (quasi) forces versus viscous forces \cite{andereck_1986,takedaNewtonian1992, Transition_2000, DutcherNewtonian2009,grossmannNewtonian2016}. The spatio-temporal fingerprints of flow transitions are commonly expressed in terms of their spectral dynamics, i.e. the characteristic temporal (frequency, $f=f(Re)$) and spatial (wavenumber, $\kappa=\kappa(Re)$) periodicities of the patterns. Thus, with increasing $Re$ the flow transitions from laminar Couette flow (LCF) to time-invariant axisymmetric counter-rotating toroidal vortices, a flow pattern known as Taylor vortex flow (TVF). While all flow patterns have one characteristic wavenumber that tends to decrease with increasing $Re$, the number of the characteristic frequencies of supercritical flow patterns can be as high as 2 and can include broadband background noise due to either successive merging and splitting of vortices or turbulent effects induced on the visualization particles within the vortices. \cite{primarysecondary_1985, andereck_1986,primarysecondary_1988,takedaNewtonian1992, DutcherNewtonian2009,Kadar2008}

Of particular recent interest in the past two decades has been understanding transition states in rheologically-complex fluids, such as viscoelastic polymer solutions. In such systems, the complex interaction between the flow field and the material structure, where the flow can modify the structure of the materials, has led to the discovery of distinctive flow patterns compared to Newtonian fluids, as well as the critical $Re$ associated to the onset of instabilities \cite{groisman2004elastic,ThomasExotic2009,liu2013polymer, Muller2013weak, Muller2013Moderate, ghanbari_2014}. In these systems, in addition to inertial and viscous forces, TC flow stability is influenced by elastic or inertio-elastic forces\cite{andereck_1986, muller1989purelyelastic,purelyelastic_1990, muller_1993_experimental, groisman_1998,groisman1998elastic,purelyelastic_1990, MullerExotic1999, Muller2013weak, Muller2013Moderate, KADAR2012TC, ghanbari_2014, Khomami2021TC}. Consequently, the dimensionless Weissenberg ($Wi$) number is introduced for expressing the ratio of elastic to the viscous response of the fluid. Upon the combination of $Wi$ and $Re$, the Elasticity ($El$) number is obtained, a parameter that signifies the ratio of elastic to inertial forces, and is expressed as \(El=Wi/Re\). Apart from aforementioned forces or time scales, an important distinction in viscoelastic fluid flows is that their viscosity functions are most commonly shear thinning. According to the scientific literature, elastic non-Newtonian fluids with negligible shear-thinning behavior exhibit altered transition sequences compared to their Newtonian counterparts \cite{muller1989purelyelastic,purelyelastic_1990,baumert1995flow, groisman_1998, MullerExotic1999, ThomasExotic2009, Muller2013Moderate}. Distinctive flow patterns thus recorded include disordered oscillations \cite{GroismanExotic1997}, oscillatory strips \cite{groisman_1998}  diwhirls \cite{KumarDiwhirls2000}, standing waves \cite{MullerExotic1999, Crumeyrolle_2002,Muller2013Moderate}, spirals and ribbons \cite{ThomasExotic2009,} and elastic turbulence \cite{liu2013polymer,Khomami2021TC}. In recent years, several works have shifted their focus to shear-thinning fluids in TC flow systems, pointing to the interplay of shear-thinning and viscoelastic response of the test fluids on the sequence and spectral dynamics of flow transitions \protect \cite{cagney2019taylor, Balabani2020TC, Balabani2021TC, laurati2009structure,calabrese2015rheology}. 

Aqueous suspensions of cellulose nanocrystals (CNCs) are a special class of viscoelastic fluids. CNCs are composed of crystalline aggregates of the polymer cellulose, as the most abundant natural polymer on earth, where they constitute rod-like nanoparticles of a few nanometers in diameter and up to several hundred nanometers in length. Remarkably, CNCs have the ability to self-assemble into liquid crystalline phases in certain structural and physicochemical conditions. Depending on the CNC concentration, several distinctions are commonly made in terms of the structure of CNC suspensions\cite{Parker2018,Kadar2021,Wojno2022}. For concentrations below the critical self-assembly concentration, such suspensions are classified as isotropic. Above the critical self-assembly concentration, the CNC suspensions consist of co-existing liquid-crystalline as well as isotropic domains, so usually referred to as biphasic. A particular feature of such birefringent suspensions is the colorful patterns that they can exhibit under polarized light\cite{Wojno2023,Wojno2022,Kadar2021,Fazilati2021,kadar2020}. The nature of the microstructural origin of birefringence as observed at flowscale depends on the assembly phase of CNCs, as observed through rheology combined with polarized light imaging, rheo-PLI, experiments\cite{Kadar2021,Wojno2022,Wojno2023}. Thus, isotropic and weakly biphasic CNC suspensions show only shear-induced birefringence with first-order interference colors \cite{ColorfulCNC2020}. With increasing CNC concentration, biphasic suspensions will transition to shear-induced orientation having higher order interference colors. As the isotropic component is diminished, the suspension will show interference colors even in quiescent conditions. Starting with biphasic suspensions having a well developed chiral nematic phase, the nature of the elastic response can be fundamentally related to various modes in which the self-assembled structures can be elastically deformed, as defined by the so-called elastic Franck constants\cite{Kuksenok1996}. However, it is known that above a critical shear rate, liquid crystalline domains can be broken into individual nanoparticles\cite{Pignon2021}. In terms of flow stability, this would be expected to effectively limit the elastic component of the suspensions. Thus, CNC suspensions constitute a niche case for analysis in TC flow for several reasons: (i) their birefringence patterns can allow for the visualization of flow patterns without the need for the addition of visualization particles; (ii) the nature of the viscoelastic response differs from that of the commonly used polymer solutions used in TC flows and (iii) the flow induced breaking of liquid crystalline domains flow make up for a case where beyond a critical $Re$ the suspensions have vanishing elastic liquid crystalline dominated material response, with likely shear-thinning remaining as the dominant factor.

To date, no systematic study has investigated the flow transitions of self-assembling suspensions in TC flow in the context of rheological and birefringence pattern visualizations. Here we study the flow stability of several biphasic CNC suspensions differing in their CNC concentration. For this, we present here a novel TC flow optical visualization setup with independently rotating cylinders that reveals the flow patterns of CNC suspensions based solely on their intrinsic birefringent properties, without the need to add flow visualization particles. The flow stability analysis is mainly based on the spatio-temporal spectral dynamics of the patterns as observed through polarized light imaging. After outlining the experimental setup and procedures, we first describe transition sequences that are representative of all the supercritical patterns observed, followed by an assessment of their stability. We then discuss the non-Newtonian effects discerned in the context of elastic versus shear-thinning effects.

\section{Materials and methods}

\subsection{Test materials}

\begin{figure}[b!]
\centering
\subfigure[]{\includegraphics[width=0.49\textwidth]{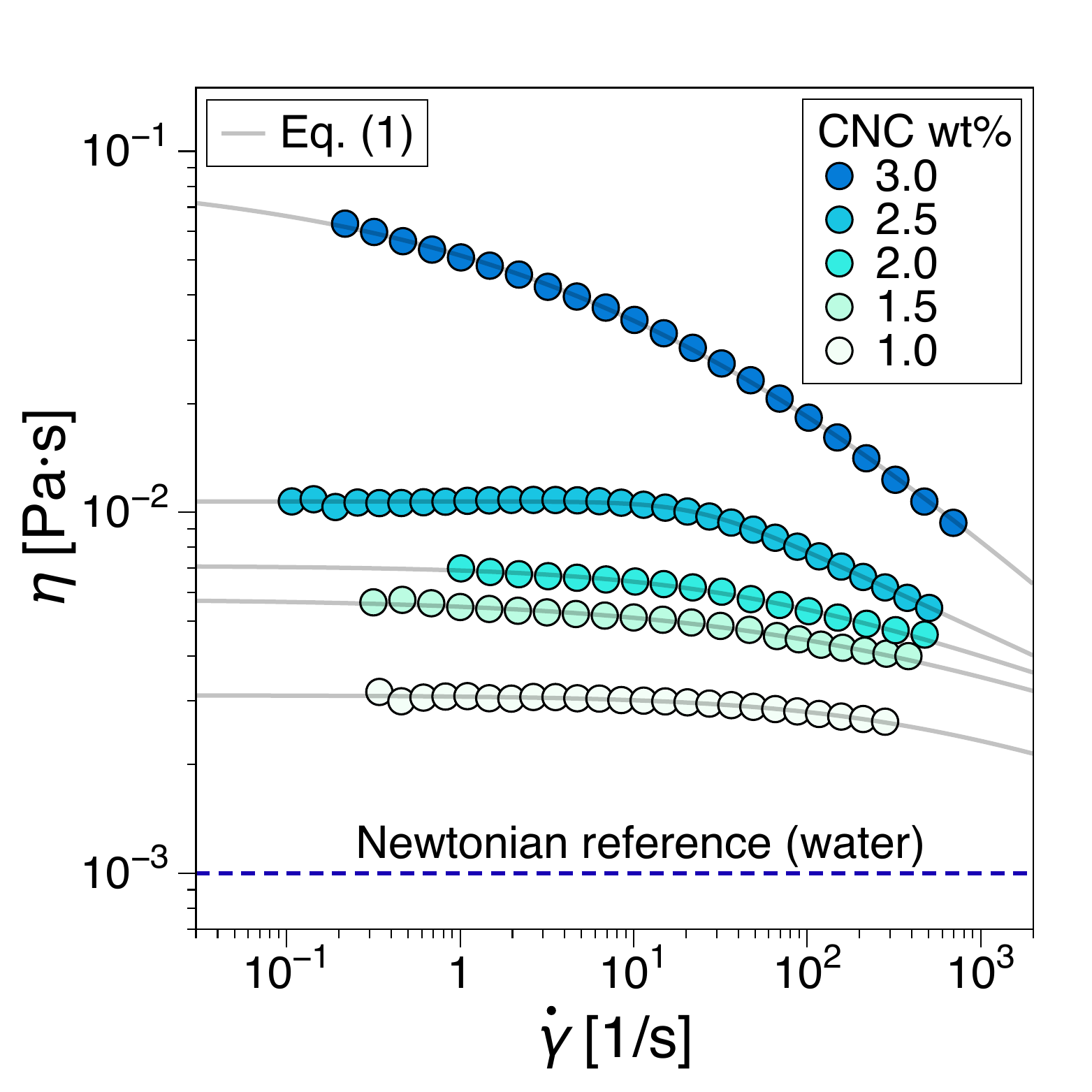}}
\subfigure[]{\includegraphics[width=0.49\textwidth]{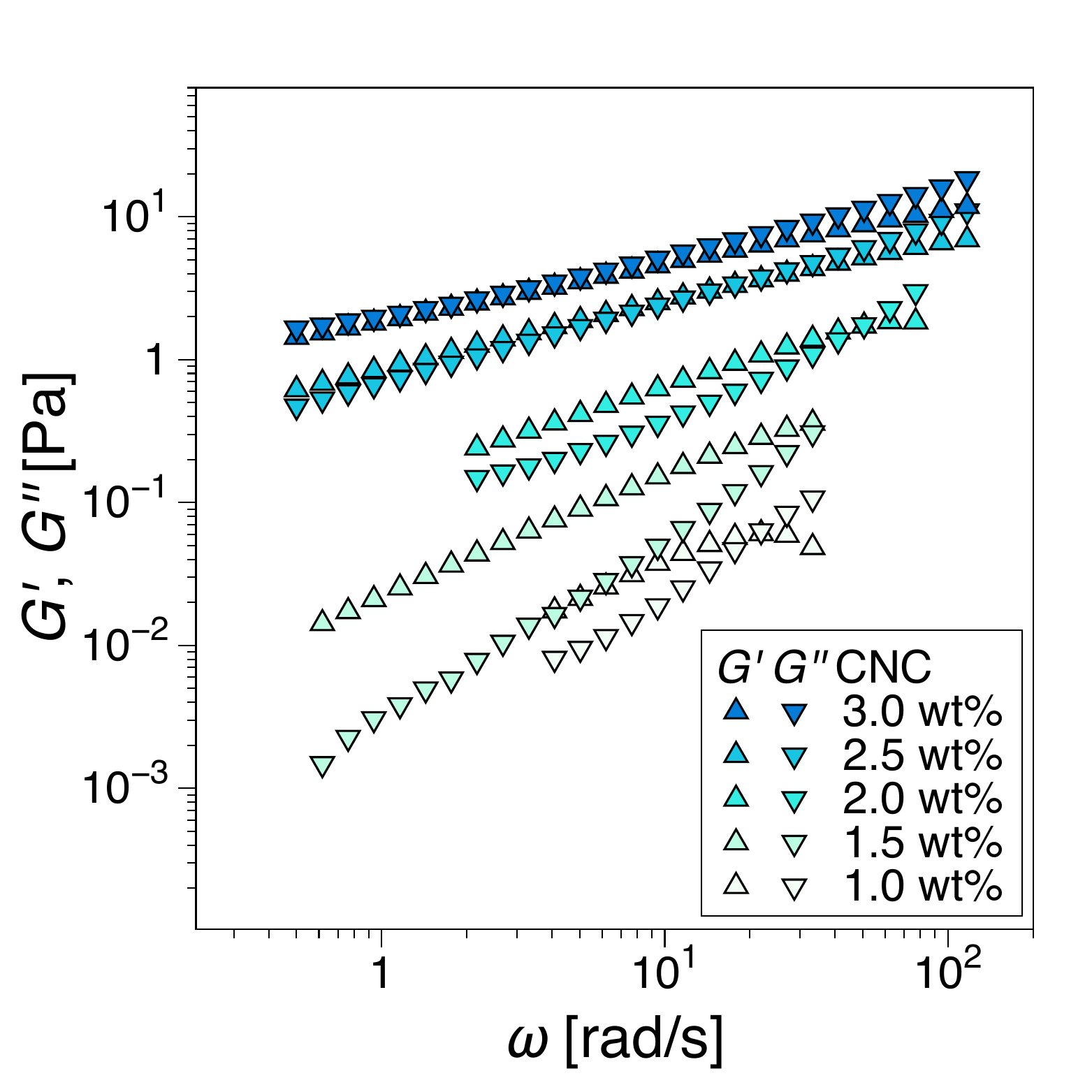}}
\caption{\label{fig1} Rheological properties of the cellulose nanocrystal (CNC) suspensions investigated: (a) shear viscosity functions from steady shear tests and (b) dynamic moduli from oscillatory shear frequency sweep tests.}
\end{figure}

The test fluids consist of distilled water mixed with visualization particles and water based CNC suspensions in five different concentrations. CNCs were purchased from CelluForce (Montreal, Canada) and were used to make CNC suspensions in Milli-Q water at 1.0, 1.5, 2.0, 2.5, and 3.0 wt \% concentrations. To this end, after mixing with Milli-Q water, the suspensions were subjected to an ultrasound bath for one hour. Subsequently, a bench shaker was used for mixing and homogenizing the suspension for 72 hours. 

The density of the suspensions was calculated based on the simple rule of mixtures:
\begin{equation}
    \rho = \rho_{\mathrm{CNC}} \phi_v + \rho_w(1-\phi_v) \label{rho}
\end{equation}
where $\phi_v$ is the volume fraction of the CNC. The resulting densities are also listed in Fig. \ref{parameters}. 

\subsection{Rheological properties}

\begin{figure}[h]
\includegraphics[width=0.48\textwidth]{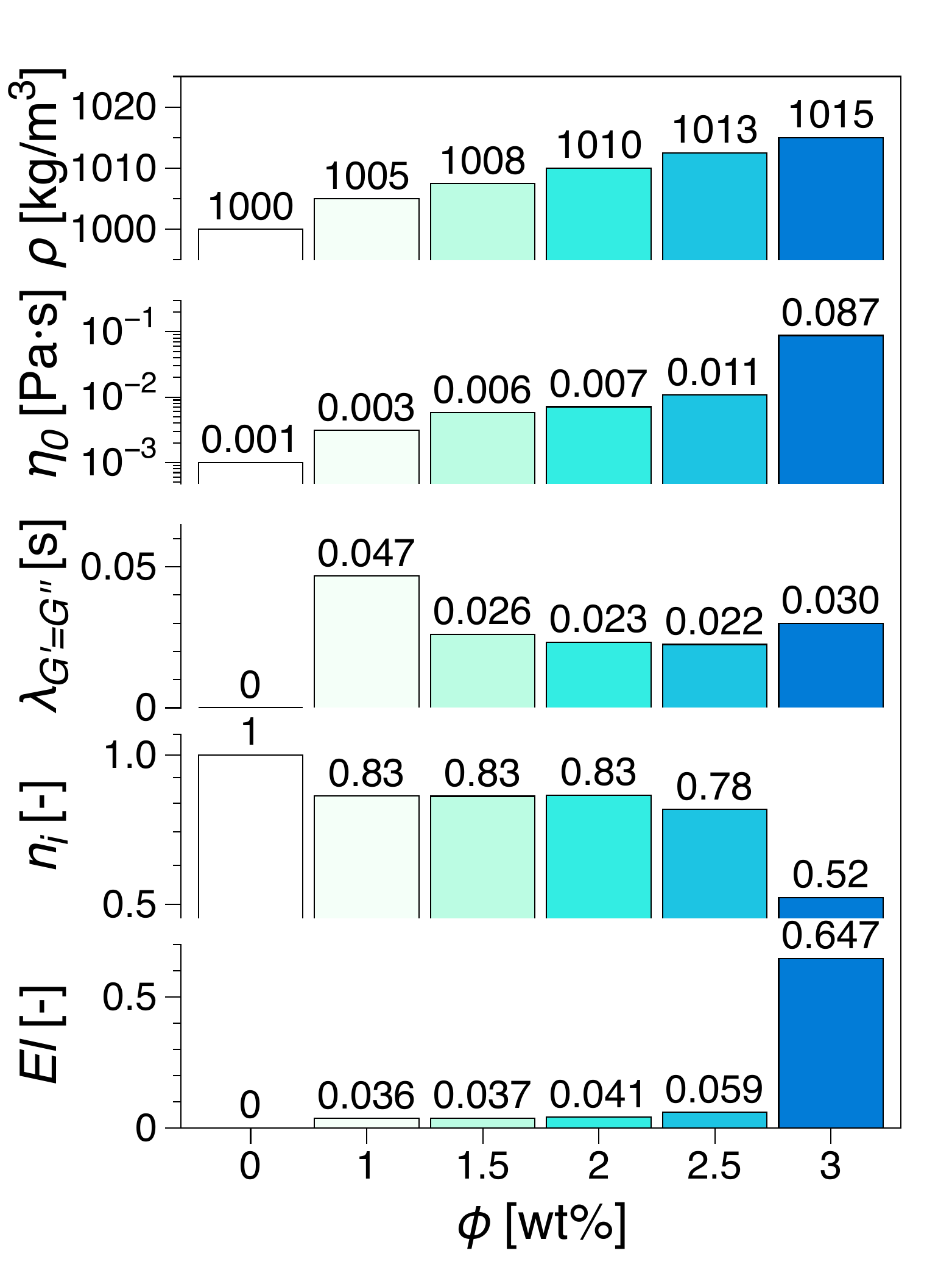}
\caption{\label{parameters} Characteristic material parameters used to define the dimensionless groups used in this study: $\eta_0$ - the zero-shear viscosity extracted from the Eq. (\ref{CY}) fits, see also Fig. SI \ref{fits_CY}, $\lambda$ - the characteristic relaxation time of the suspensions, \(\lambda_r = 1/\omega_{G''=G'} \) in Fig. \ref{fig1}.}
\end{figure}

Steady and oscillatory shear tests were performed on an Anton Paar MCR702e Space (Graz, Austria) rotational rheometer in single motor-transducer configuration using double-gap and bob-cup measuring geometries. Fig. \ref{fig1} represents the steady shear viscosity functions and angular frequency dependent dynamic moduli of the test samples at room temperature. The viscosity functions in Fig. \ref{fig1}(a) were fitted with the Carreau–Yasuda model:
\begin{equation}
 \eta (\dot{\gamma}) =  \eta_0 ( 1 + (\lambda_{CY}\dot{\gamma})^\alpha)^\frac{n_i-1}{\alpha} \label{CY}   
\end{equation}
where ${\eta_0}$ and $\eta_\infty$ stand for viscosity as the shear rate tends to zero and to infinity, respectively; $\lambda_{cy}$ and $n_i$ denote the model relaxation time and shear-thinning index (flow index), respectively; $\alpha$ is a parameter describing the transition from the Newtonian plateau to the shear-thinning region. The fit parameters are plotted in Fig. \ref{fits_CY} in the supplementary information. Linear viscoelastic dynamic moduli from frequency sweep tests performed at a constant strain amplitude of 70, 20, 4, 4, and 1 \% for 1.0, 1.5, 2, 2.5, and 3.0 wt\%, respectively, are shown in Fig. \ref{fig1}(b). All concentrations show a liquid-like behavior in the terminal region, $G''>G'$, with the highest concentration potentially approaching a gel-like material response. We note that the linear viscoelastic dynamic moduli essentially characterize the linear (Newtonian-plateau) region of the viscosity functions in Fig. \ref{fig1}(a). The longest relaxation times of the terminal region, i.e. the inverse of the angular frequency at which this cross-over occurs \(\lambda = 1/\omega_{G''=G'} \), are also represented in Fig. \ref{parameters}.

Based on previous results with similar preparation methods\cite{Wojno2022} and polarized light microscopy analysis, $\phi_{CNC} \in (1,2]$ wt\% are expected to be isotropic suspensions, whereas $ \phi_{CNC} > 2$ are expected to be biphasic. Since 2 wt\% CNC shows a uniform background, we can consider it isotropic, Fig. \ref{POM}(a) while 3 wt\% appears biphasic, see Fig. \ref{POM}(b), with~10 $\mathrm{\mu}$m sized agglomerates.

\subsubsection{Rheo-SAXS}

To further gain a better understanding of the microstructural dynamics of the CNC suspensions, rheological tests were performed simultaneously with small-angle x-ray scattering experiments, rheo-SAXS at the CoSAXS beamline \cite{RheoSAXS2012} , Max IV Laboratory, Lund University, Sweden. A standard SAXS polycarbonate cup-bub geometry was used in single motor-transducer configuration. The measuring geometry had an inner cylinder radius of 24.5 mm with a 25 mm outer cup radius (radius ratio: 0.98). The high radius ratio of the geometry ($\approx 0.96$) ensures that the flow was laminar up to the maximum shear rate tested, 100 $^{-1}$s. Furthermore, we have also visually confirmed the absence of instabilities (data not shown). To quantify the flow microstructure, we present here only Hermans orientation parameter, $\left< P_2 \right>$, from azimuthal integration within $q \in [1.4,2.4]\cdot 10^{-2}$ Å$^{-1}$, where $q$ is the magnitude of the scattering vector. The scattering experiment was set up such that the incident X-rays pass through the axis of the concentric cylinders around the middle of the flow column. Thus, the scattering experiments are representative of the velocity - vorticity directions (so-called (1-3) plane). By approximating the orientation distribution function with a Legendre series expansion in $\cos \varphi$ of the orientation distribution function containing only even terms, $I\varphi) \approx \sum_{n=0}^6 a_n P_{2n}(\cos \varphi)$, $\left< P_2 \right>$ can be defined as:
\begin{equation}
\left< P_2 \right> = \frac{\int_0^\pi \frac{1}{2} \left( 3 \cos^2\varphi - 1 \right) I(\varphi) \sin \varphi d\varphi }{\int_0^\pi I(\varphi) \sin \varphi d\varphi}
\end{equation}
where $\varphi$ is the azimuthal angle of the scattering pattern and $I(\phi)$ is the scattering intensity within the integrated $q$-range. Within the integration limits used, $\left< P_2 \right> \in [-0.5,1]$, where $\left< P_2 \right> = 1$ signifies that all CNCs are oriented in the perpendicular direction to the flow, and $\left< P_2 \right> = -0.5$ indicate that all CNCs are oriented in the flow direction, and $\left< P_2 \right> = 0$ corresponds to random orientation. Due to the low CNC concentrations employed, only the 3 wt\% sample is presented.

\subsection{Experimental setup}

\begin{figure}[h]
\subfigure[]{\includegraphics[height=6cm]{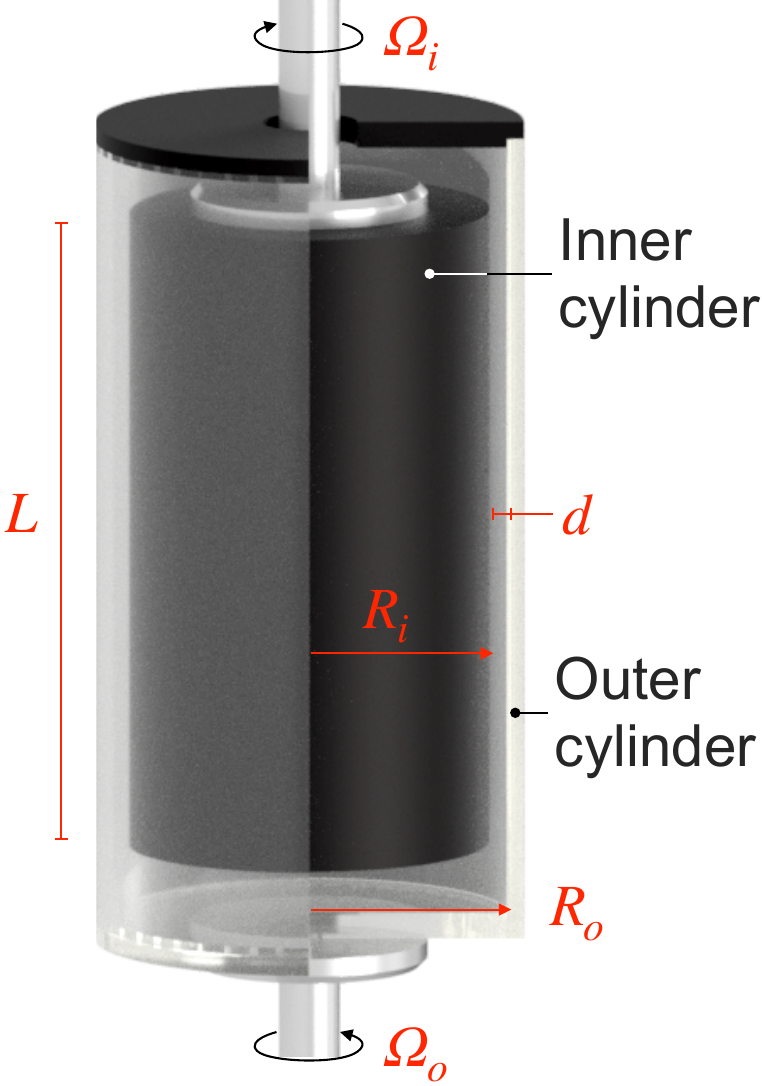}}
\quad
\subfigure[]{\includegraphics[height=6cm]{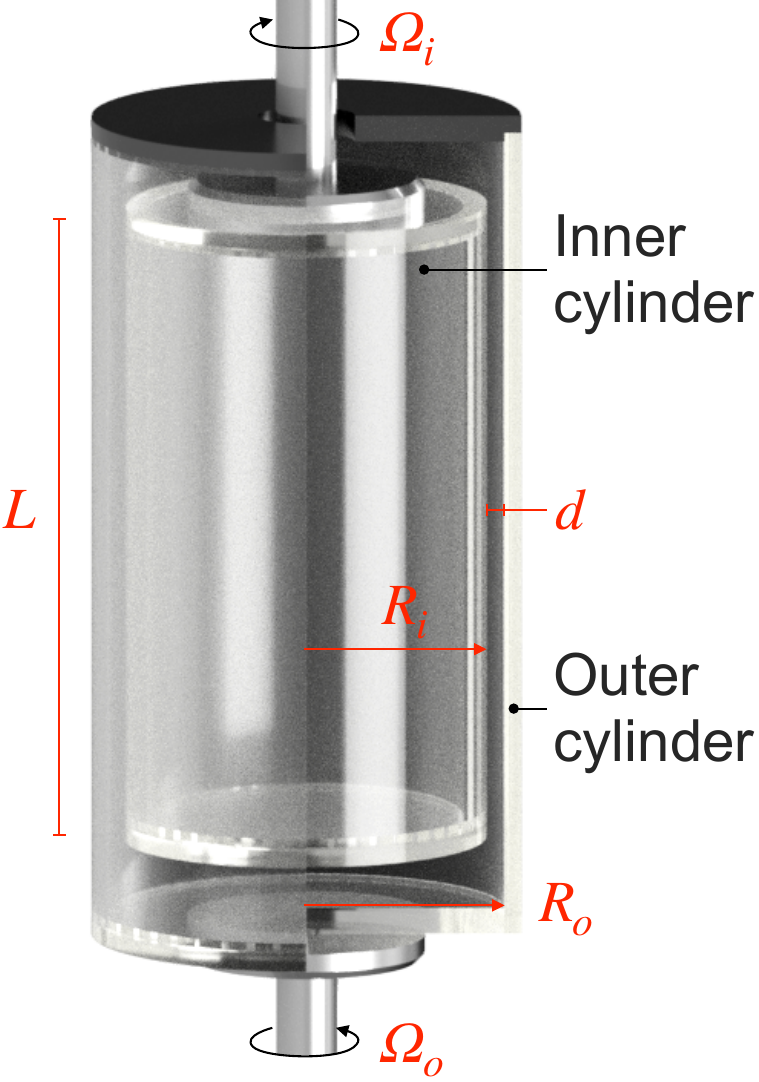}}
\caption{\label{fig2} The two independently-rotating Taylor-Couette (TC) flow geometries used in the present study: (a) TC cell for visualization with reflective particles consisting of a transparent glass outer cylinder and an aluminum inner cylinder, for the reference sample (water+visualization particles) and (b) TC cell for cross-polarized light visualizations consisting of a transparent glass outer cylinder and translucent polycarbonate inner cylinder. The two are exact dimensional replicas of each other.}
\label{_Fig1}
\end{figure}

The flow stability experiments were performed using two custom-design Taylor-Couette (TC) visualization flow cells, Fig. \ref{_Fig1}. The TC flow cells were mounted on the same rheometer used for the rheological characterization albeit in a separate motor-transducer configuration. The inner and outer cylinders have radii of $Re_i=20.5$ mm and $Re_o=22.5$ mm, respectively, giving a geometry radius ratio of \(\epsilon = R_i/R_o =  0.91\). The height of flow column is $L=68.7$ mm, with a resulting aspect ratio of \(\Gamma = L / d = 34.4\). All tests were run at the temperature of 23 $^{\circ}$C (ambient temperature). The rotation of the inner and/or outer cylinder was controlled by the upper and lower motors of the rheometer. 

\subsection{Dimensionless numbers}

To account for the relative rotation of the inner (${\Omega_i}$) and outer (${\Omega_o}$) cylinders, respectively, we define
\begin{equation}
    \beta = \frac{\Omega_o}{\Omega_i+|\Omega_o|} \label{beta}
\end{equation}
where $\Omega_{tot} = \Omega_i+ \left| \Omega_o \right|$ is the total net angular speed of the flow cell. Thus, $\beta = 0$ corresponds to a rotating inner cylinder while the outer one is at rest; for $\beta = -1$ the outer cylinder is rotating and the inner one is stationary. The minus sign thus signifies counter-rotation. Five flow cases have been thus considered: $\beta = 0, -0.25,-0.5,-0.75$ and $-1$. The Reynolds number for each cylinder can be defined as 
\begin{equation}
Re^{(i,o)} = \frac{\rho(\phi) \cdot R_{i,o} \cdot |\mathit{\Omega}_{i,o}| \cdot d_{\beta}}{\eta_0(\phi)} \label{Re}
\end{equation}
where $d_\beta$ is the relative gap between the two concentric cylinders that depends on $\beta$
\begin{align}
d_\beta &= 
  \begin{cases}
    d \cdot (1-|\beta|)  & \text{if }  \beta=0 \; \text{ i.e.} \; Re = Re^{(i)}\\
    d \cdot |\beta| & \text{if } \beta <0 \; \text{ i.e. } \; Re \neq Re^{(i)}  . \label{delta_beta}
  \end{cases}
\end{align}
 It is thus convenient to report results using a 'total' Reynolds number defined as 
 \begin{equation}
     Re = Re^{(i)} + Re^{(o)}.
 \end{equation}
Thus, for any $\{Re,\beta\}$ reported, the corresponding instantaneous $Re$ of the inner and outer cylinders are therefore $Re^{(i)} = Re(1-|\beta|)$ and $Re^{(o)} = Re|\beta|$. We briefly note that in Eq. (\ref{Re}) $\eta_0$ is the zero-shear viscosity as extrapolated from Eq. (\ref{CY}), in contrast to some previously published works where they use a shear rate dependent viscosity for shear-thinning fluids \cite{lacassagne2020vortex,caton2006, Balabani2020TC}. For a more comprehensive notation, the critical $Re$ numbers are expressed as $Re^{\beta,\phi_{\mathrm{CNC}}}_{cr}$.  

In the case of Non-Newtonian fluids, as explained in the introductory section, the Weissenberg number $Wi$ is introduced to account for the relaxation timescale of the materials. In the current work, $Wi^{(i)}$ and $Wi^{(o)}$ correspond to the inner and outer cylinders, respectively 
\begin{equation}
Wi^{(i,o)} = \frac{\lambda \cdot R_{i,o}}{|\mathit{\Omega}_{i,o}| \cdot d_{\beta}} \label{Wi}
\end{equation}
By combining (the total) $Re$ and $Wi$, the Elasticity number, $El$, is defined as 
\begin{equation}
    El = \frac{Wi}{Re} = \frac{\lambda \eta}{\rho d^2}
\end{equation}
which is essentially a function only of fluid properties and flow geometry. Based on the $El$ range in Fig. \ref{parameters} we can classify suspensions with $\phi_{CNC} < 3$ wt\% as weakly elastic, $El << 1$, and $\phi_{CNC} = 3$  as moderately elastic, $El \approx 1$\cite{Muller2013Moderate}. 

From a different perspective, the flow of liquid crystals can be characterised by the Ericksen number, $Er$, as the balance between viscous and elastic forces. In contrast to $Wi$, the characteristic material parameters are defined in the framework of the Leslie-Ericksen theory and specific Frank elastic constants as $Er^{(i,o)} = \frac{\gamma_1 v d}{K}$ 
where $\gamma_1$ is the rotational viscosity as defined by the Leslie-Ericksen viscosities\cite{Leslie1968,Ericksen1961}, $\gamma_1=\alpha_3-\alpha_2$ (the $\alpha_2$ and $\alpha_3$ are Leslie viscosity coefficients), $v^{(i,o)}$ is the velocity of the inner/outer cylinders and $K$ is a Franck elastic constant.
We note here that determining experimentally the Leslie-Ericksen viscosities remains a challenge\cite{RobertOrr2011} and the theory has been applied mainly to nematogenic liquid crystals whereas CNCs form a chiral nematic phase. The elastic Franck constants has been investigated for nematic and cholesteric tactoids in biphasic CNC suspensions by Bagnani et al.\cite{mezzenga2021elastic} while estimates of Leslie-Eriksen viscosities for CNCs can be found in the work of Noroozi et al.\cite{Noroozi2014}. However, considering that the low concentrations considered in the study, even if present, the mesophase is expected to break up at relatively low shear rates\cite{Pignon2021,Noroozi2014}, all three Frank elastic constants would tend to zero, making $Er$ a unsuitable measure of flow stability.

\subsection{Experimental protocol}

The flow stability analysis was performed using ramp-up speed tests starting at a rotational velocity of zero and linearly increasing it up to the maximum capability of the rheometer.  The ramp rate was chosen such that for any flow case the total rotational speed ramp is constant for all CNC suspension flows, i.e. $\forall \beta, d \Omega_{tot} / dt = 0.022$. A criterion for meeting quasi-steady-state conditions in a ramped-up TC flow was defined as $dRe/dt^* < 1$ by Dutcher and Muller \cite{DutcherNewtonian2009} on Newtonian fluids, where \(t^*\) stands for a dimensionless time via the viscous time scale \(t_\nu = (\rho d^2/\eta)\). Using the same principle, the ramp rates in the present work correspond to \(dRe/dt^* \leq 0.16 \) for the CNC suspensions. A special test protocol was chosen for the Newtonian reference (water) and was performed at \(dRe/dt^* \leq 7.29 \), as a compromise between creating optimal quasi-steady-state conditions and the difficulty in handling large amounts of video data. We briefly note that for the Newtonian reference $\beta=0$ case, the experimental critical Reynolds number for the onset of instabilities was $Re_{cr1} \approx 210$. This is considerably higher than theoretical predictions based on linear stability analysis \cite{Chandrasekhar,EsserGrossmann}, where $Re_{cr1}|_{\epsilon = 0.91} \approx 136$.

\subsection{Flow visualization and spectral analysis}

\begin{figure}[h!]
\centering
\includegraphics[width=\textwidth]{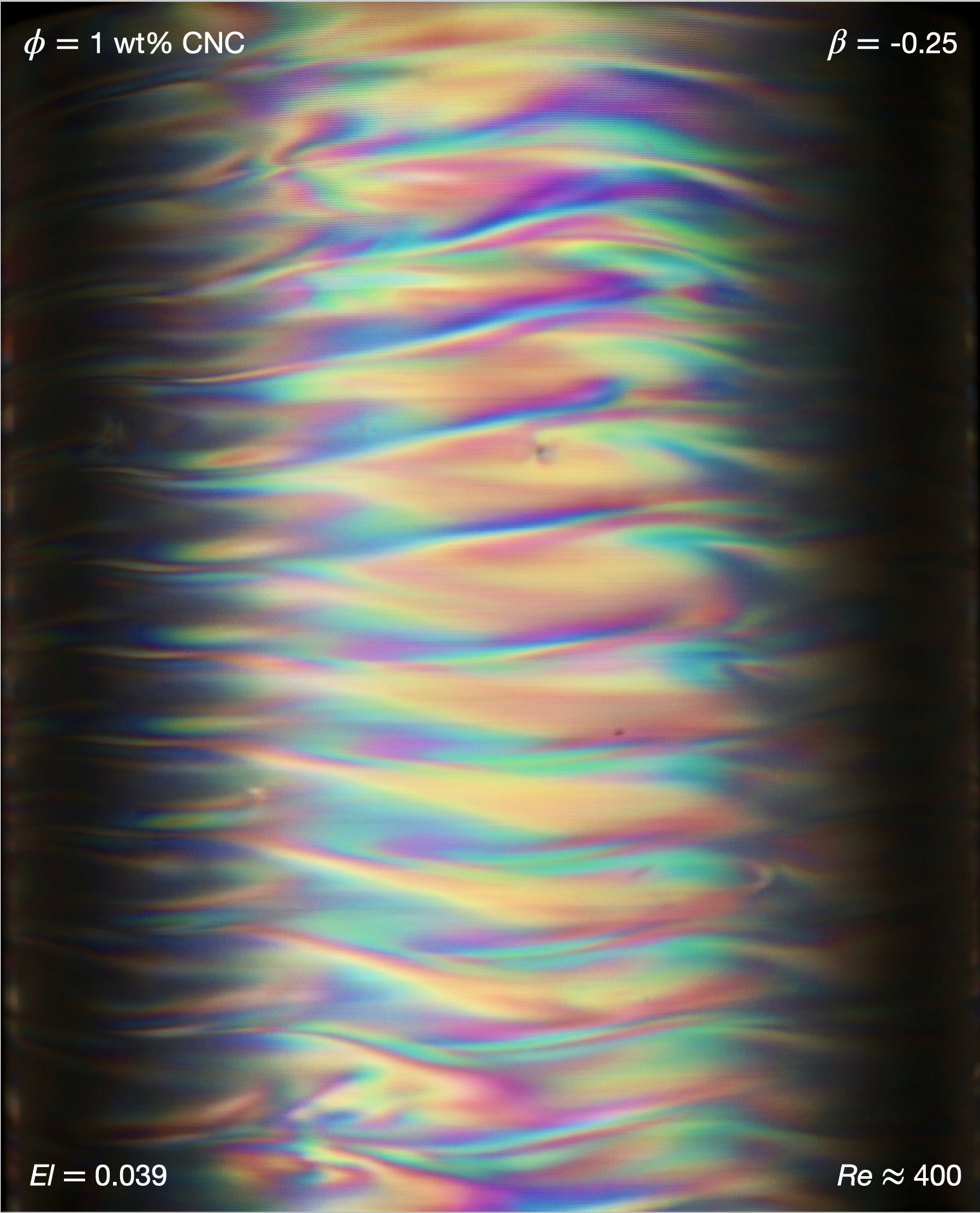}
\caption{Example of supercritical Taylor-Couette flow visualization of CNC suspensions ($\phi = 3$ wt\% CNC, $\beta = -0.25$, $\left | Re_{tot} \right | \approx 400 $). The colors are natural as visualized directly from the birefringent properties of the suspensions using a cross-polarized light optical visualization system.} \label{fig4}
\end{figure}

One of the experimental features of the present work distinguishing it from previous studies is the design of a customized TC flow visualization setup for cross-polarized optical imaging (PLI). The visualization setup comprised a Canon 90D DSLR camera (Tokyo, Japan) equipped with a Canon L-series 100 mm macro lens. LED studio lights were used as light source. Standard TC flow visualization setups operate in reflection mode, meaning the light source is placed on the same side of the geometry with the camera and light reflected by the visualization particles is captured by the camera. In contrast, the PLI setup we developed operates in transmission mode, with the concentric cylinders placed between the light source and the camera together with two linear polarizers oriented at 90$^\circ$ relative to each other placed between the flow cell and camera and light source, respectively. Due to fluid induced orientation during manufacturing, the translucent polycarbonate cylinder exhibits weak birefringence, however, when tested on water (non-birefringent) no flow patterns could be distinguished. In contrast, using the liquid crystalline test samples, beautifully colored TC flow patterns could be readily observed, see Fig. \ref{fig4}.

During the ramped experiments, HD video (1920 \(\times\) 1080 px) recordings of the flow patterns were performed at 100 fps. Space-time diagrams were then produced by extracting a vertical line of pixels passing through the middle of the flow domain from from the video frames that were added successively to create a new image (~45000 \(\times\) 1080 px), as it is common in most similar studies. Thus, in the new image the y-axis corresponds to the the height of the flow column and the x-axis to time and correspondingly $Re$. Subsequently, 2D Fourier transform (FFT) was applied using a moving window procedure. The window size was 500 px and the window increment 1 px. For each window $j$ the 2D Fourier transform of the grayscale intensity function $g_j(z,t)$ is
\begin{equation}
    G_j(z,t) = \int_{-\infty}^{+\infty}g_j(z,t)e^{-i(\kappa)z,2 \pi f \cdot t} dz dt
\end{equation}
where $z$ is the coordinate along the axis, and $(\kappa, f)$ are the characteristic wavenumber and frequency of the window analyzed. Thus, based on the number of characteristic $(\kappa,f)$ and their dynamics, specific TC flow patterns can be identified.

\section{Results and discussion}

\begin{figure}[t]
\includegraphics[width=\textwidth]{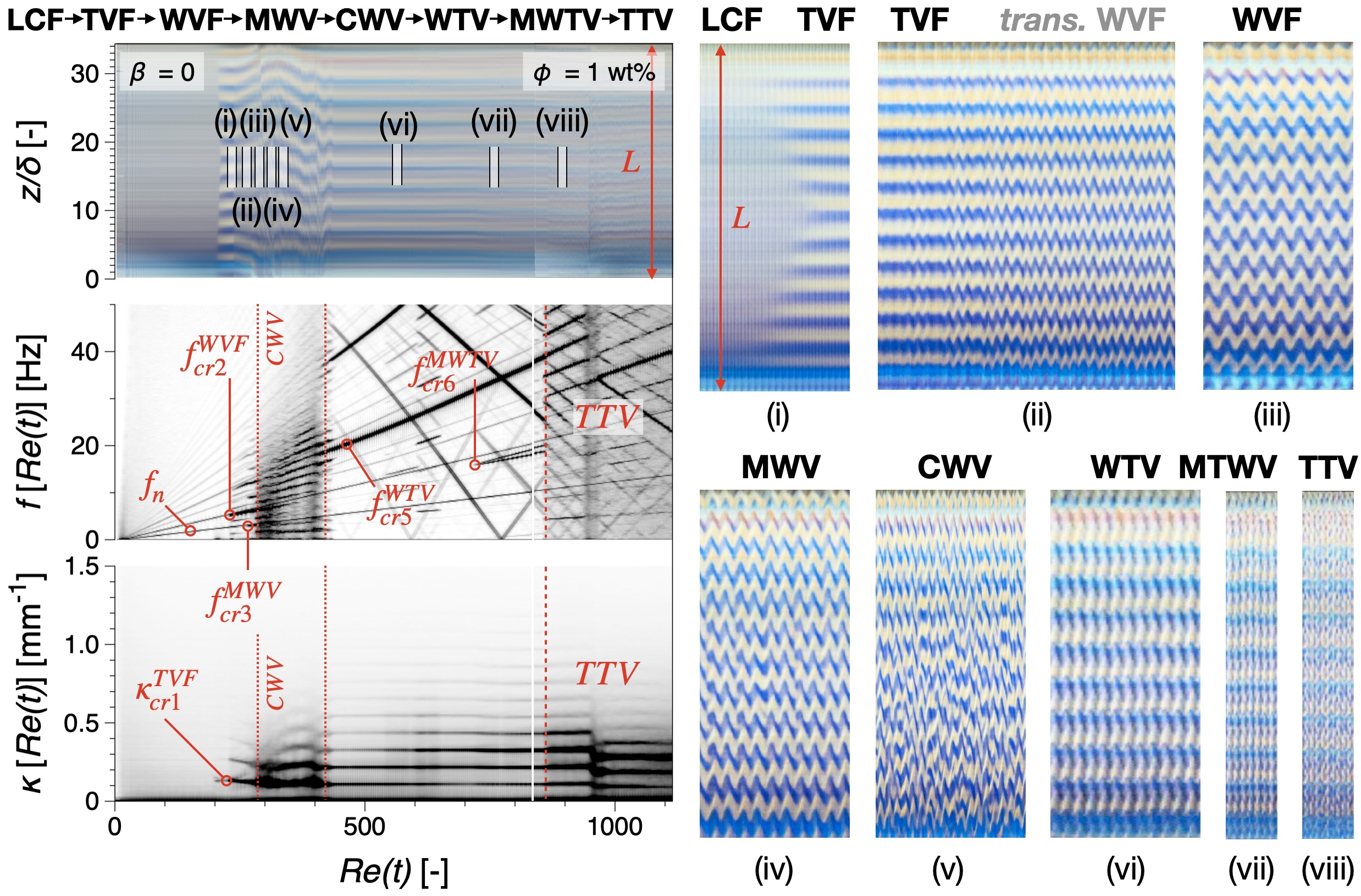}
\caption{\label{fig5} Transition sequence for the 1 wt\% CNC suspension, $\beta=0$. The left column compares the space-time visualization (top), temporal spectrogram (middle) and spatial spectrogram (bottom), while on the right side specific instability modes are highlighted as details from the space-time visualization.}
\end{figure}
We first examine only the unique transition sequences for $\phi = 1$ wt\% CNC as function of $\beta$ since they essentially contain nearly the entire spectrum of instability modes identified in this study. Thereafter we discuss the influence of elasticity on the flow stability and pattern characteristics. Subsequently, the results are discussed in light of the interplay between the elasticity, shear-thinning and the structural evolution of CNC suspensions.

\subsection{Characteristic instability modes and the influence of relative cylinder rotation}

\begin{figure}[h!]
\subfigure[]{\includegraphics[width=0.9\textwidth]{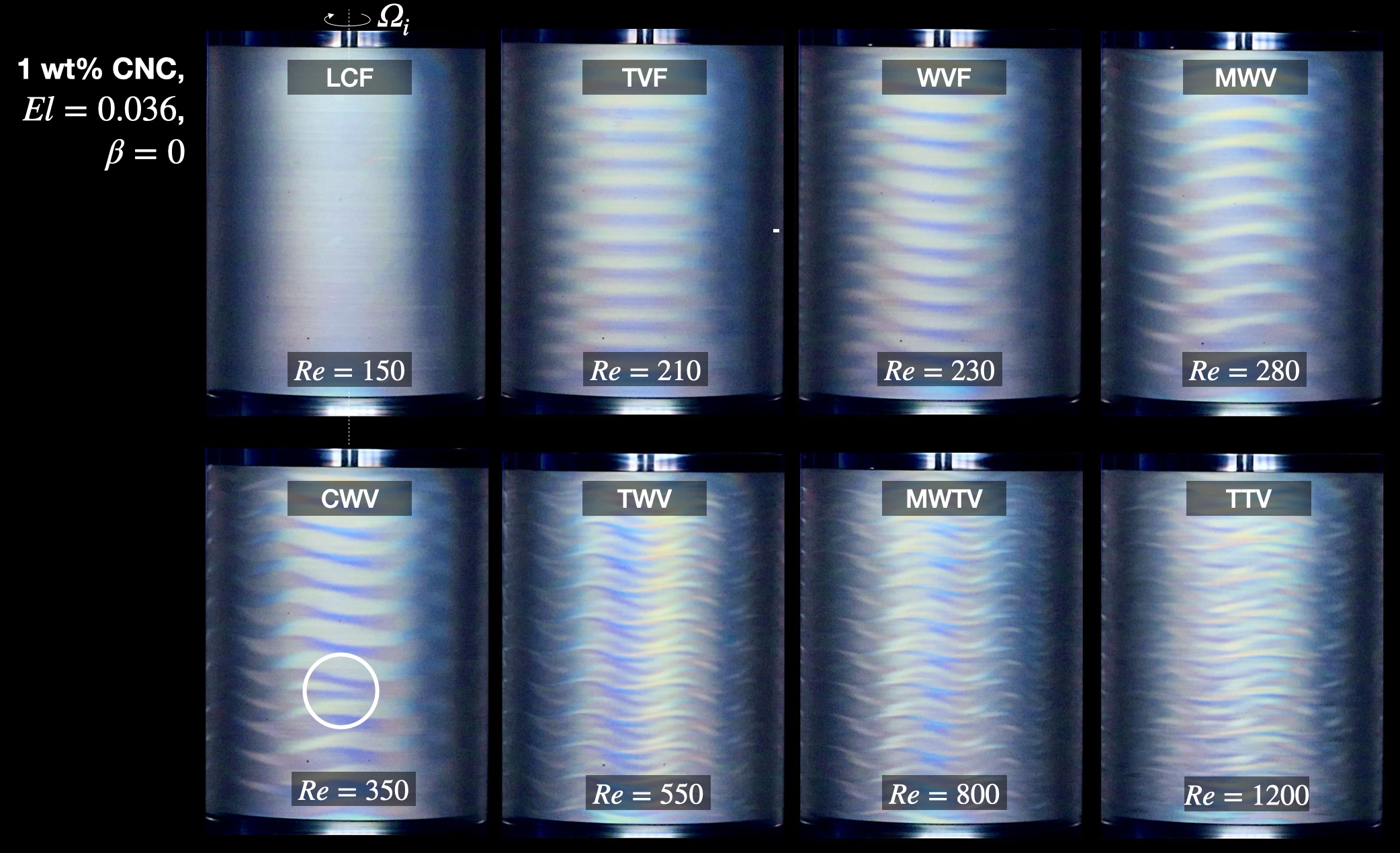}}
\subfigure[]{\includegraphics[width=0.9\textwidth]{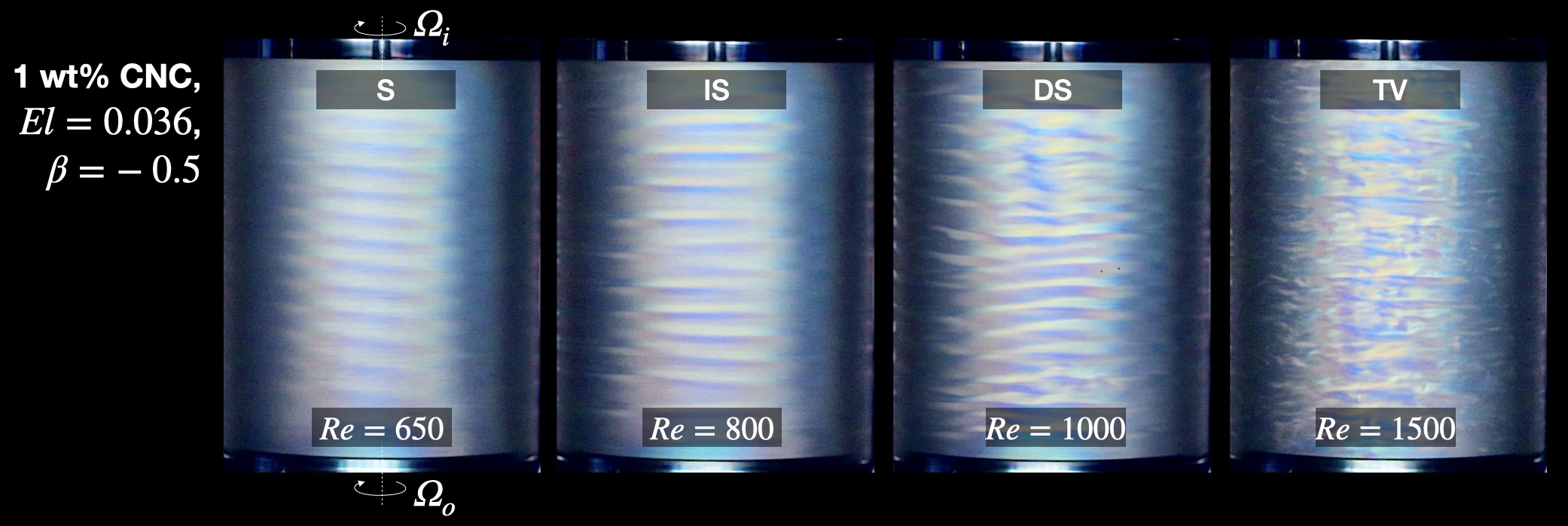}}
\subfigure[]{\includegraphics[width=0.9\textwidth]{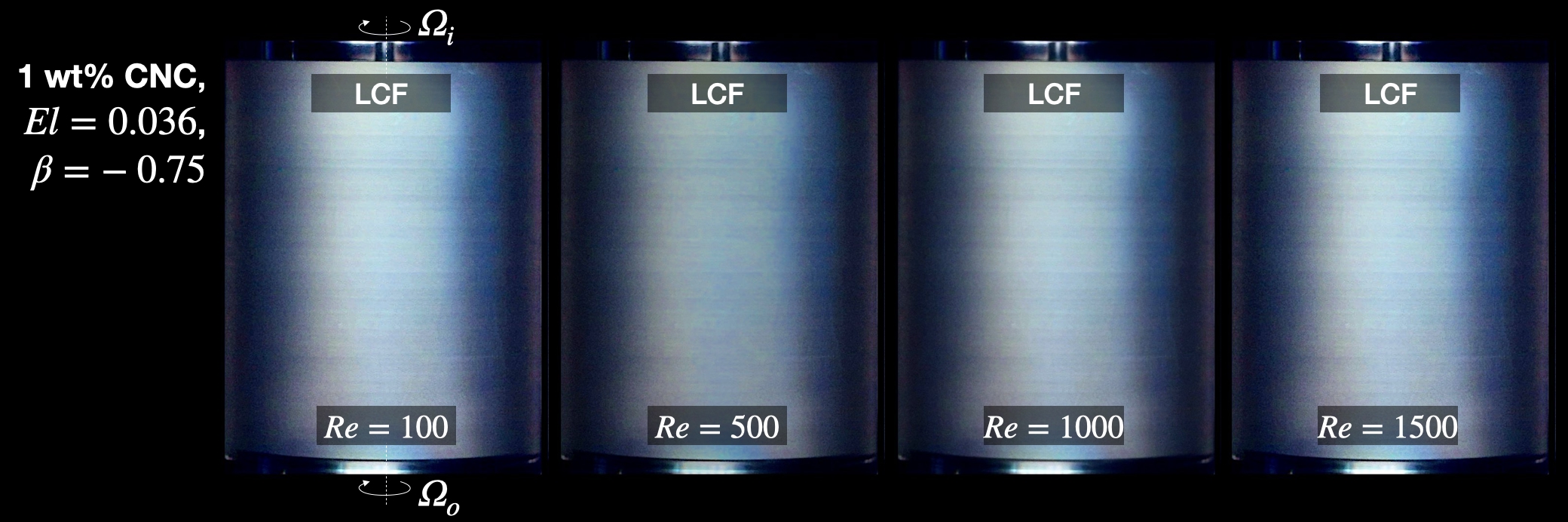}}
\caption{\label{fig6} Flow visualizations at selected $Re$ showcasing the main instability modes observed. The visualizations correspond to the data presented in Fig. \ref{fig5}-\ref{fig8}: (a) $\beta = 0$, (b) $\beta = -0.5$ and (c) $\beta = -0.75$. In (a) the collapse of a pair of vortices is highlighted for $Re = 350$.}
\end{figure}

\begin{figure}[t]
\includegraphics[width=\textwidth]{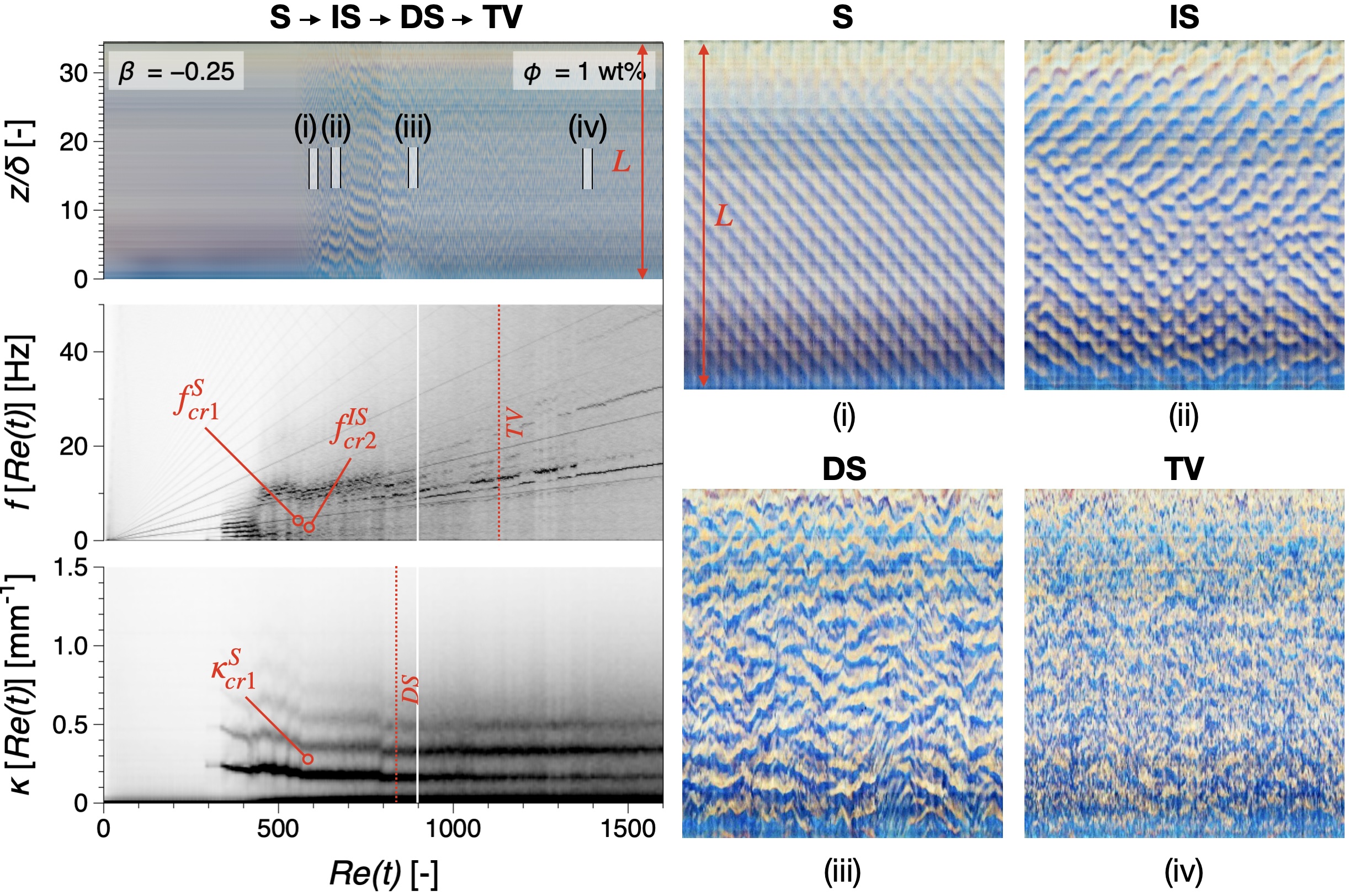}
\caption{\label{fig7} Transition sequences for the 1 wt\% CNC suspension $\beta = -0.5$. The left column contains the space-time visualization (top), temporal spectrogram (middle) and spatial spectrogram (bottom), while on the right side specific instability modes are highlighted as details from the space-time visualization.}
\end{figure}

The flow transitions for 1 wt\% CNC at three different relative cylinder rotations, i.e., $\beta = 0, -0.5$ and $-0.75$, are represented in Fig. \ref{fig5}, \ref{fig6}, and \ref{fig7}, respectively. The figures compare the space-time diagrams, and details therefrom, with scalar plots of the (temporal) frequency spectra and (spatial) wavenumber spectra as function of $Re$. Complementary still frame extracts from the video recordings representative of the patterns identified are summarized in  Fig. \ref{figAll2}. We note that in all temporal spectrograms the rotational frequency of the rotating cylinders, $f_n^{(i,o)}$ has the highest signal-to-noise ratio and a significant number of higher harmonics present in the spectra. This can be readily identified as constant ramp peaks that span the entire test duration. As a reference, the instability modes for the Newtonian case (water) are presented in Fig. \ref{figB1a} in Appendix \ref{A}.

%Corresponding torque diagrams are summarized in Fig. \ref{fig6} (and in the SI Fig.***). 
\begin{figure}[t]
\includegraphics[width=\textwidth]{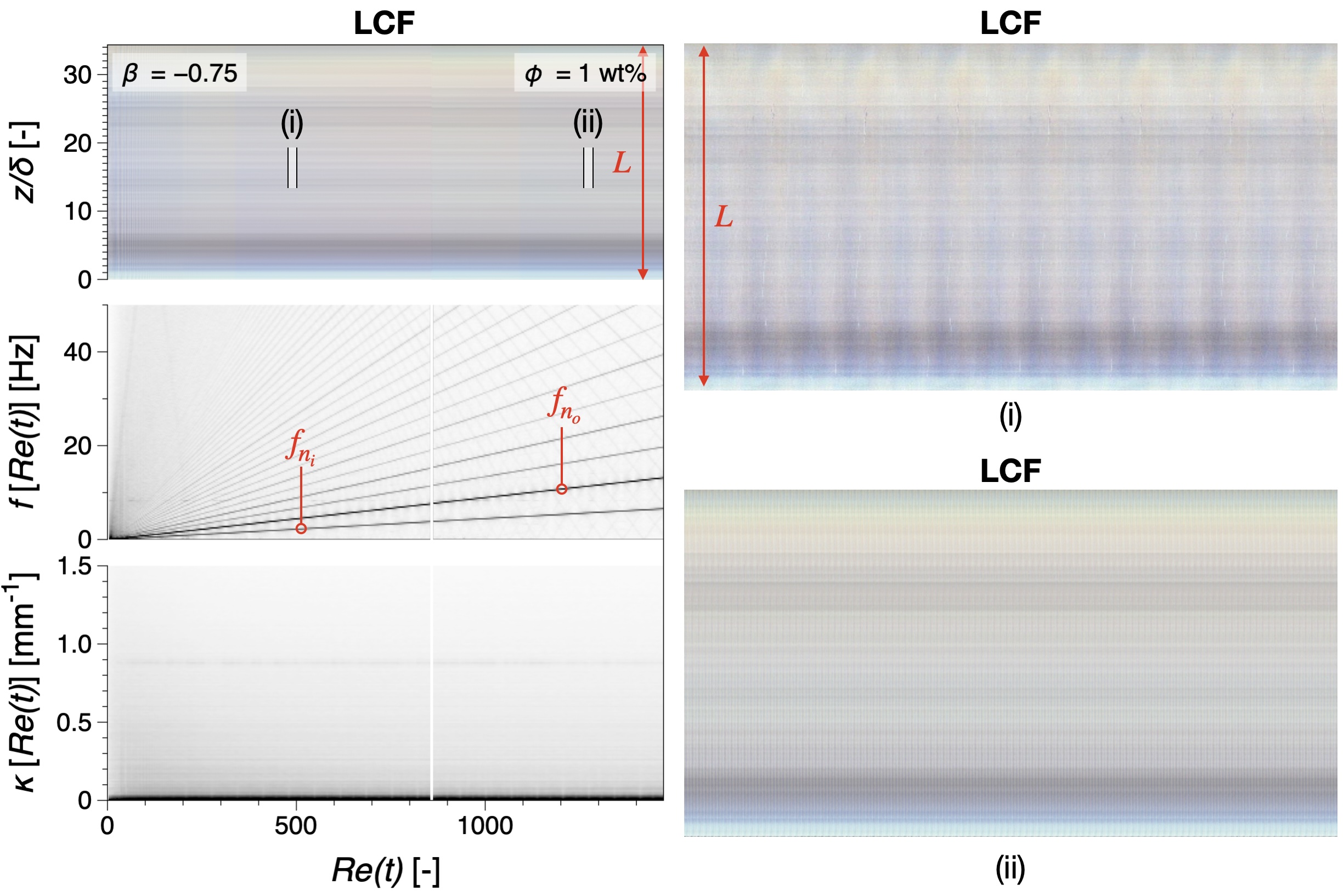}
\caption{\label{fig8} Transition sequence for the 1 wt\% CNC suspension, $\beta=-0.75$. The left column contains the space-time visualization (top), temporal spectrogram (middle) and spatial spectrogram (bottom), while on the right side specific instability modes are highlighted as details from the space-time visualization.}
\end{figure}

Based on the data in Fig. \ref{fig5}-\ref{fig8}, the instability modes are identified, and their sequence with increasing $Re$ are essentially Newtonian-like\cite{taylor1923,andereck_1986,coles1965transition}. 

For $\beta = 0$, Fig. \ref{fig5}, the Laminar Couette flow (LCF) cascades into the axisymmetric Taylor vortex flow (TVF, $Re^{0,1}_{cr1} \approx 193$), where parallel colored stripes appear in the space-time diagram, see Fig. \ref{fig5}(i), characterized only by one wavenumber $\kappa^{TVF}$, Fig. \ref{fig5}. Taylor vortices become axially unstable and evolve into wavy vortex flow (WVF, $Re^{0,1}_{cr2} \approx 229$), where colorful waves traveling in the azimuthal direction can be readily identified. We note that with the appearance of azimuthal waves, there is a stabilization region, Fig. \ref{fig5}(ii), before the waves settle for WVF, see Fig. \ref{fig5}(iii). The onset of the wavy regime is identifiable in terms of spectral dynamics by the onset of a characteristic (temporal) frequency of the waves, $f^{WVF}$, while the axial motion of the secondary flows typically leads to a decrease in $\kappa$. At higher $Re$ the wavy regime undergoes a wave modulation in frequency, Fig. \ref{fig5}(iv), resulting in the modulated wavy vortices (MWF, $Re^{0,1}_{cr3} \approx 260$). In the temporal spectrogram, the characteristic pattern frequencies bifurcate with the addition of the low modulation frequency component, $f^{MWF}$. The onset of the chaotic wavy regime (CWV, $Re_{cr4} \approx 283$) features a broadband background noise and multiple characteristic temporal frequencies. This includes a broadening of $\kappa$\ in the spatial spectrogram. These spectral characteristics are due to the rapid successive collapsing and splitting of vortices that characterize the pattern, see Fig. \ref{fig5}(v). The following supercritical instability mode is expected to be wavy turbulent vortices (WTV, $Re_{cr4}^{0,1} \approx 420$). Interestingly, at the observation scale the spectra does not capture the broadband characteristics expected for material particles in turbulent flow. This is further confirmed by the similar spectra obtained for the Newtonian reference case, see Fig. \ref{figB1a}. This in contrast to CWV where the broadband noise present in the spectra corresponds to events occurring at flow scale. Instead, WTV is characterized by the onset of $f^{WTV}$ while $\kappa$ maintains a well defined fundamental and higher harmonics. A modulation frequency could be detected with increasing $Re$ marked as a distinct pattern here, namely modulated wavy turbulent vortices (MWTV, $Re_{cr6}^{0,1} \approx 700$). Ultimately, the flow cascades into the turbulent Taylor vortex regime (TTV, $Re_{cr7}^{0,1} \approx 870$) with a wide spectrum of frequencies and broadband noise. To summarize the full spectrum of instabilities identified for CNC suspensions in the $\beta = 0$ case follows a first Newtonian sequence (Sequence 1) of instabilities: LCF $\rightarrow$ TVF $\rightarrow$ WVF $\rightarrow$ MWV $\rightarrow$ CWV $\rightarrow$ WTV $\rightarrow$  MWTV $\rightarrow$ TTV.

The $\beta = -0.25$ counter-rotation case exhibits a similar sequence in instabilities to $\beta = 0$. However, new flow modes emerge with $\beta = -0.5$, where both cylinders counter-rotate at equal speeds, Fig.\ref{fig7}. We briefly note that in counter rotation mode the flow field is separated into two zones by the zero-velocity plane, thus enabling the onset of instabilities in both zones. Consequently, based on the simple visualization method employed we cannot fully resolve the spatio-temporal topology of the instability modes. Following LCF, the first stability mode is a non-axisymmetric spiral flow (S, $Re_{cr1}^{-0.25,1} \approx 614$) instability mode, characterized by both one (temporal) frequency and (spatial) wavenumber $f^S, \kappa^S$\ corresponding to the azimuthal and axial periodicity of the spiral flow, respectively, Fig. \ref{fig7}(i). The spiral mode translates into interpenetrating spirals (IS; $Re_{cr2}^{-0.25,1} \approx 633$), Fig. \ref{fig7}(ii). IS appears to have a transient characteristic with increasing $Re_{tot}$ with initially one new low-frequency component, $f_{\# 1}^{IS}$ apparent in the spectra, followed by noise in that frequency range and the addition of a new lower frequency component(unmarked) and the decay of $f_{\# 2}^{IS}$. Gradually, all identifiable pattern peaks broaden, and there is a broadband background noise in both the temporal and spatial spectrogram, and this signals the onset of disordered spirals (DS, $Re_{cr3}^{-0.25,1} \approx 900-1000$), Fig. \ref{fig7}(iii). Eventually, the background noise in both $f$ and $\kappa$ increases to a point where clear spatio-temporal patterns are difficult to identify with the flow entering a new regime we call turbulent vortices (TV, $Re_{cr4}^{-0.25,1} \approx 1200-1500$), Fig. \ref{fig7}(iv). To summarize, the full spectrum of instabilities identified for CNC suspensions in the $\beta = -0.5$ case also follows a second Newtonian sequence (Sequence 2) of instabilities: LCF $\rightarrow$ S $\rightarrow$ IS $\rightarrow$ DS $\rightarrow$ TV.

For all counter-ration cases with $\beta \leq -0.75$ the LCF flow is stable at all $Re$ as exemplified in Fig. \ref{fig8}. 

\begin{figure}[t]
\includegraphics[width=0.65 \textwidth]{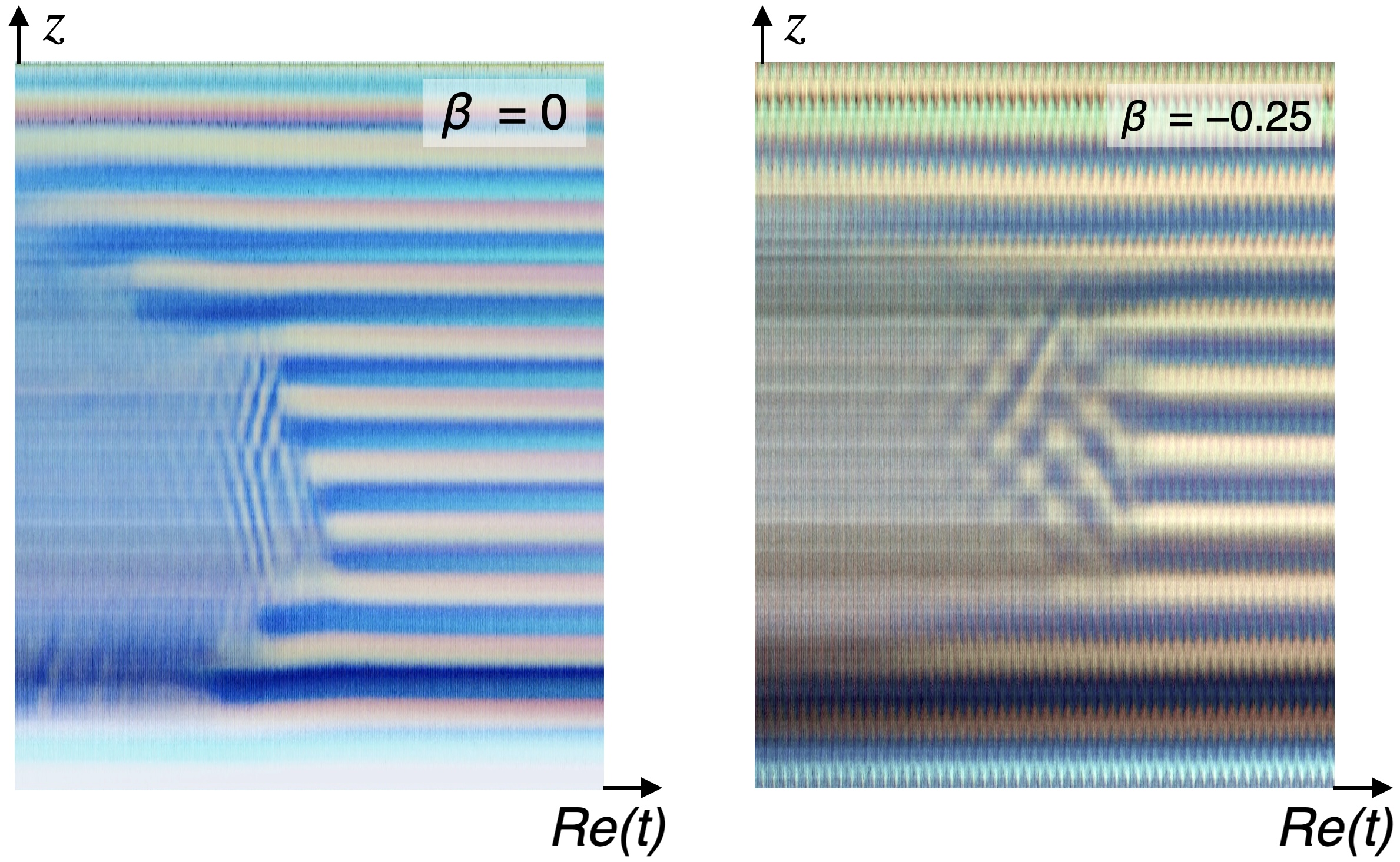}
\caption{\label{fig9} Examples of pre-TVF instabilities in the form of ribbon-like spirals observed in 3 wt\% CNC, $\beta = 0,-0.25$.}
\end{figure}

One exception from the Newtonian patterns described is found for 3 wt\% CNC in the form of ribbon-like spirals instabilities just before the onset of TVF for $\beta=0$  and $-0.25$, see Fig. \ref{fig9}.

\subsection{Non-Newtonian effects on pattern formation and stability}

\begin{figure}[p!]
\subfigure[]{\includegraphics[width=0.49\textwidth]{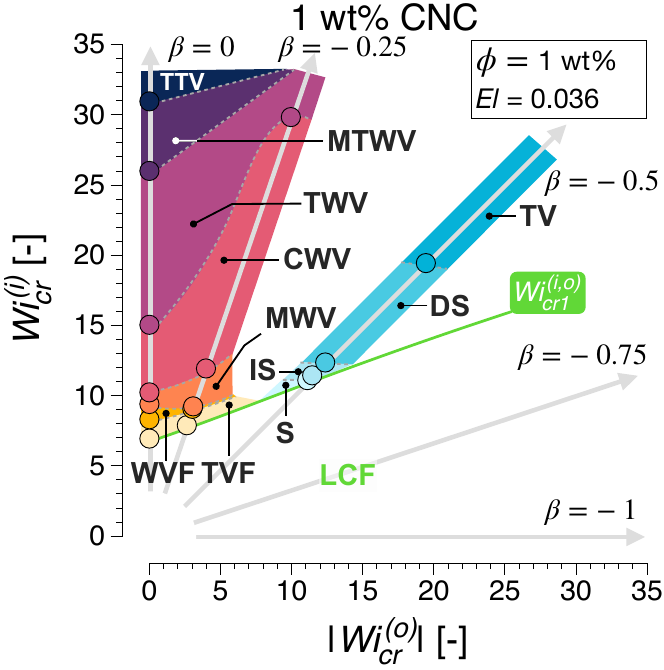}}
\subfigure[]{\includegraphics[width=0.49\textwidth]{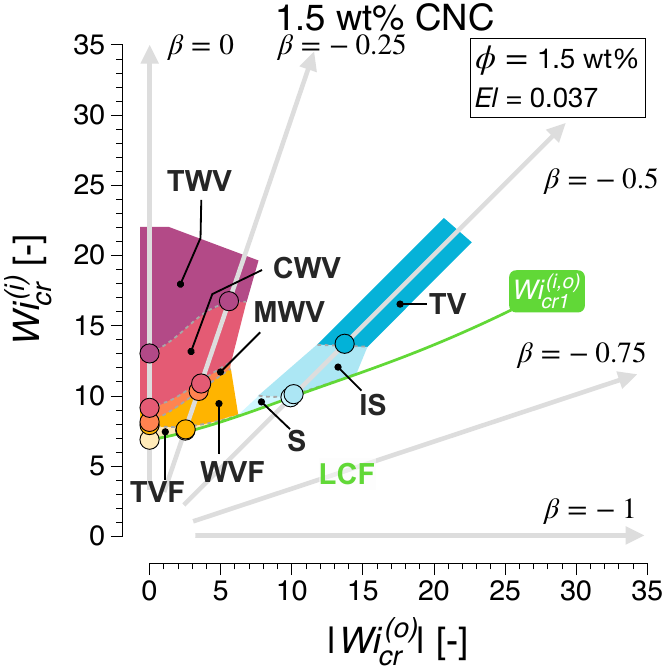}}
\subfigure[]{\includegraphics[width=0.49\textwidth]{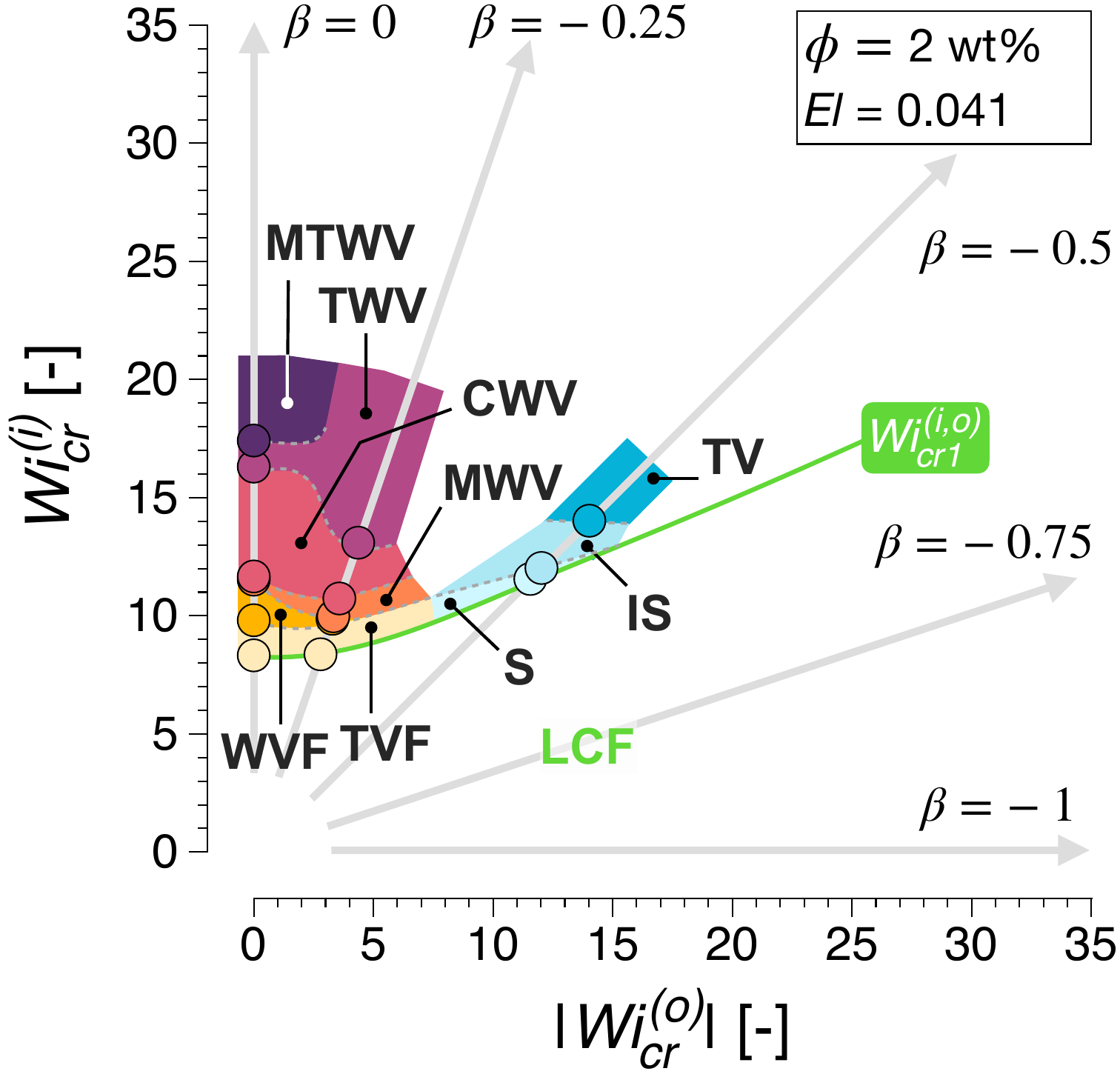}}
\subfigure[]{\includegraphics[width=0.49\textwidth]{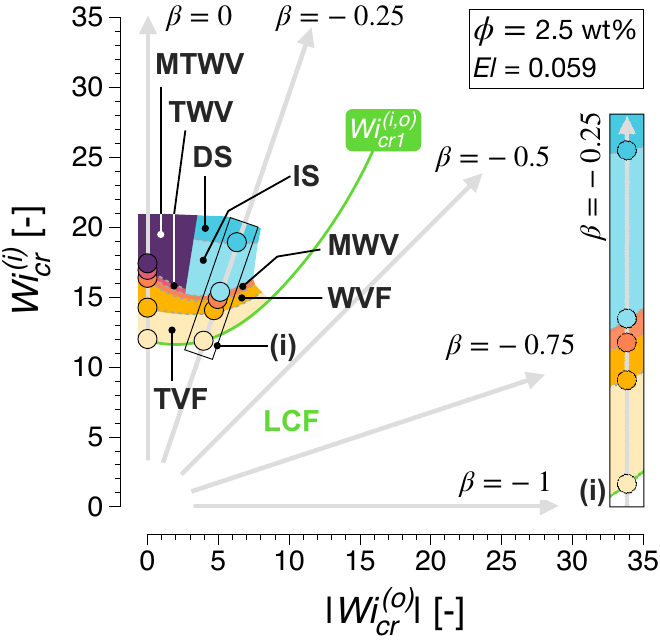}}
\subfigure[]{\includegraphics[width=0.49\textwidth]{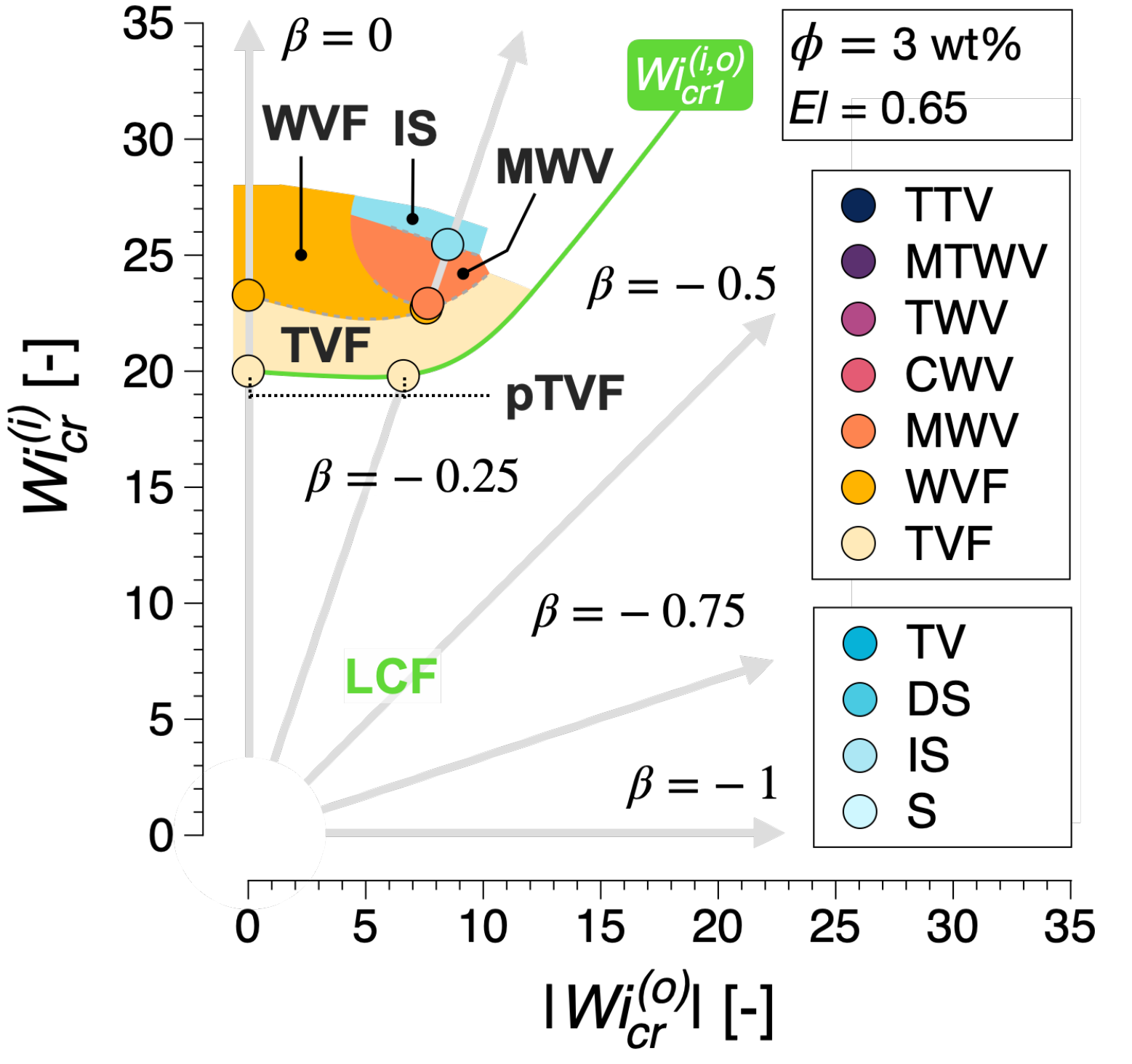}}
\subfigure[]{\includegraphics[width=0.49\textwidth]{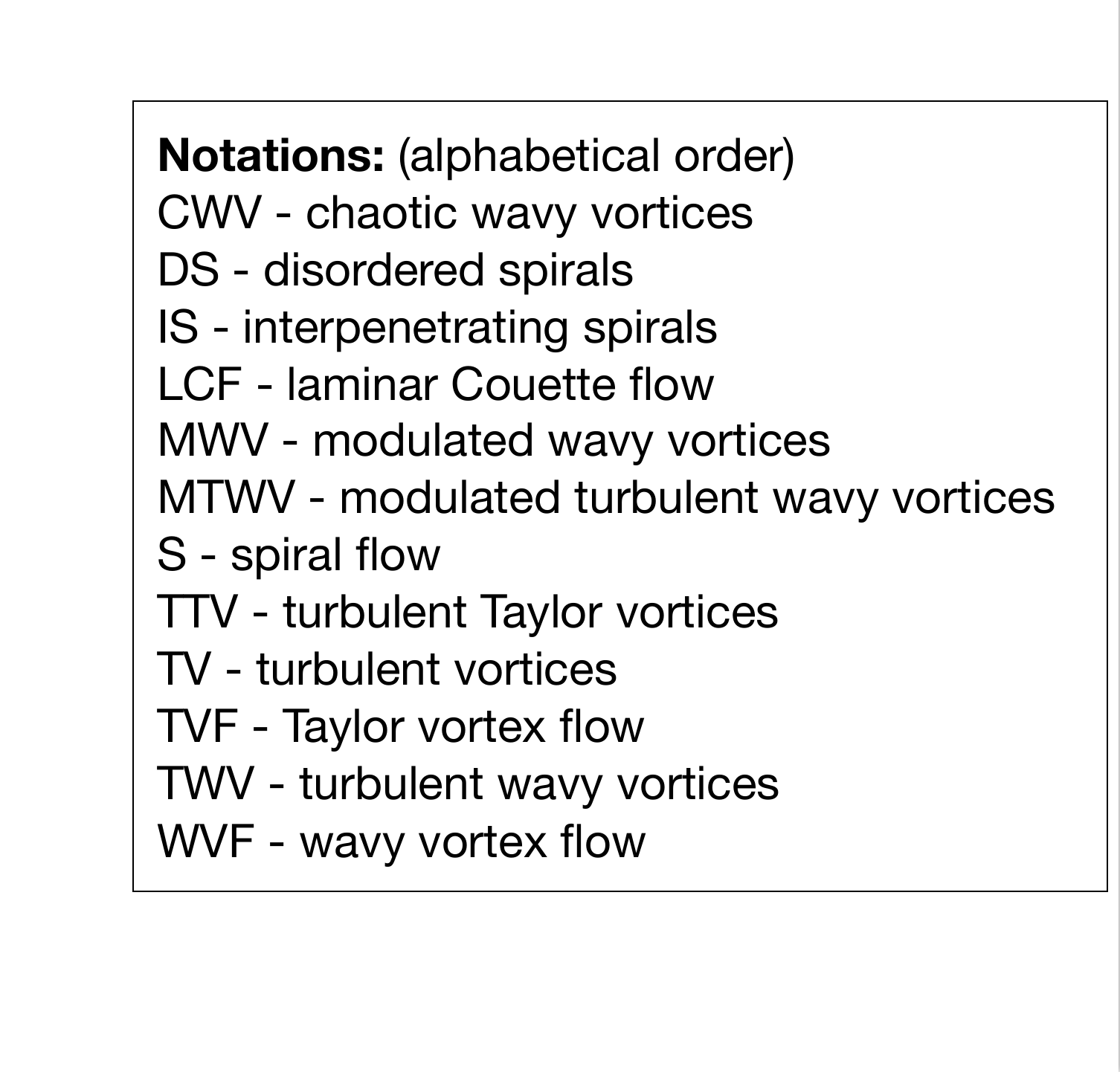}}
\caption{\label{fig10} Stability diagrams in the form of $Wi_{cr}^{i}$  vs. $|Wi_{cr}^{o}|$ for all CNC suspensions, $\phi = $(a) 1, (b) 1.5, (c) 2, (d) 2.5, (e) 3 wt\% CNC. A summary of notations is listed in (f).}
\end{figure}

Here we focus on identifying non-Newtonian effects on pattern formation and flow stability. Fig. \ref{fig10} summarizes the transition sequences recorded for all $\beta$ and CNC concentrations investigated. The stability diagrams therein map the characteristic flow patterns in a $Wi_{cr}^{i}$ vs. $|Wi_{cr}^{o}|$ plot. Thus, $\beta = 0$ corresponds to $|Wi_{cr}^{o}| = 0$ and thus a vertical line, and $\beta = -1$ corresponds to $|Wi_{cr}^{i}| = 0$ , a horizontal line and so on. Overall and as previously mentioned, nearly the entire spectrum of supercritical flow cascades can be mapped using Newtonian-like instability modes. The maximum $Wi$ attainable for each concentration is limited by the increase in viscosity, and consequently the number of accessible states dwindle with increasing $El$, e.g. compare Fig. \ref{fig10}(a) and (e). However, non-Newtonian effects could still be distinguished in (i) the characteristic wavenumbers of the instabilities $\kappa = \kappa(\phi_{CNC})$ as well as pre-TVF instabilities at the highest concentration; (ii) the $Wi$-range over which some instability modes are stable, (iii) a transition sequence that starts with flow patterns characteristic of Sequence 1 but then transitions into modes characteristic of Sequence 2 and (iv) the stabilization / destabilization of critical instability mode parameters with increasing CNC concentration, $\phi_{CNC}$.

\subsubsection{Pattern characteristics}

\begin{figure}[h!]
\includegraphics[width=\textwidth]{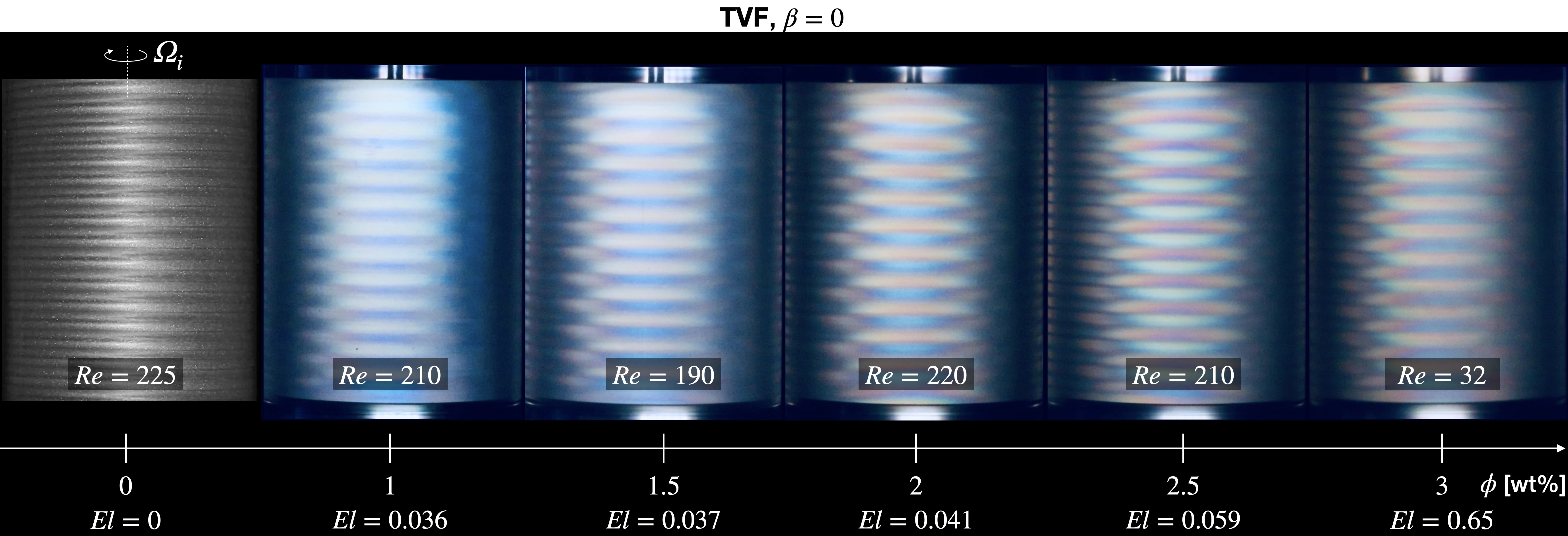}
\caption{\label{fig11} Influence of CNC concentration on the Taylor vortex flow (TVF) instability mode ($\beta = 0)$. Note that the flow visualizations correspond in most cases to slightly supercritical conditions, $Re \geq Re_{cr1}$.}
\end{figure}

A visual comparison between the TVF flow patterns as a function of $\phi_{CNC}$ is presented in Fig. \ref{fig11}. For the Newtonian case, the size of a Taylor vortex pair in TVF has been shown using linear stability analysis to be twice the gap between the concentric cylinders, $\Delta z = 2 d$\cite{taylor1923}, where $\Delta z$ is the average size of a pair of vortices in the flow column. Deviations from this have been associated with non-Newtonian fluids. Shear thinning and elasticity have been shown to have opposite effects thereon. While patterns with $\Delta z > 2 d$ (decrease in $\kappa$) have been shown to occur due to the effect of shear thinning within the flow field\cite{bahrani_2015_shearthinning,beavers1974tall, Escudier1995taylor,lange2001vortex,cagney2019taylor,cagney2020taylor, alibenyahia2012revisiting},  the results for elastic non-shear thinning fluids such as Boger fluids reveal $\Delta z < 2 d$ (increase in $\kappa$)\cite{larson1992, purelyelastic_1990}.  

This can be readily observed in the present data as well, by simply comparing the number of vortices present in the column for TVF, Fig. \ref{fig11}. Thus, for $\phi_{CNC} = 0$, 34 vortices in the fluid column roughly correspond as expected to half the height of the flow cell, i.e. $\kappa \approx \pi/d$. However, for $\phi_{CNC} > 0$ wt\% $\kappa^{TVF} << \pi/d$, with the total number of Taylor vortices in the column being 28 ($\kappa^{TVF} \approx 1.28$) for $\phi_{CNC} = 1$ wt\% and down to only 21 ($\kappa^{TVF}= 0.96$) for $\phi_{CNC} = 3$ wt\%. Overall, a decrease in $\kappa$ for CNC suspensions compared to the Newtonian case can be readily observed across the entire spectrum of instability modes identified, e.g. compare, for $\beta = 0$, $\kappa = \kappa(Re)$ for the Newtonian reference in Fig. \ref{figB1a} to the Non-Newtonian case of 1 wt\% case in Fig. \ref{fig5}. 

It needs to be emphasized that $\kappa(Re)$ is also significantly different between the Newtonian case and the CNC suspensions. For the latter, $\kappa|_{\phi_{\mathrm{CNC}}=ct.}$ does not decrease significantly with increasing $Re$. This is also distinct from $\kappa$ dynamics observed in similar shear thinning and elastically dominated transition sequences.\cite{larson1992, purelyelastic_1990,cagney2020taylor}.

\subsubsection{Transition from axisymmetric and azimuthally-periodic patterns to asymmetric modes}

As identified in the previous section, two distinct transition sequences (Sequences 1 and 2) are present for $\phi_{CNC} \leq 2$, Fig. \ref{fig10}. The first sequence starting with TVF (Sequence 1) occurs for $\beta \geq -0.25$ while the second sequence starting with S (Sequence 2) is limited to $\beta = -0.5$. In contrast, for $\phi_{CNC} \geq 2.5$, the two transition sequences co-exist for $\beta = -0.25$, Fig. \ref{fig10}(d) and (e). Initially TVF is followed by WVF and MWV, however, higher order transitions are destabilized by the relative counter-rotation of the outer cylinder, and the CWV regime was not detected for $\beta \neq 0$.

\subsubsection{Onset of instabilities and their $Re$ range}

Increasing $\phi_{CNC}$ has a significant effect on the range of $Wi^{(i,o)}$ for almost all flow patterns. Taking $\beta = 0$ (|$Wi_{o} = 0$|) and for $\phi_{CNC} = 2$ wt\%, it can be observed that the TWV region has been partially suppressed compared to the low concentration cases.

\begin{figure}[h!]
\subfigure[]{\includegraphics[width=0.49\textwidth]{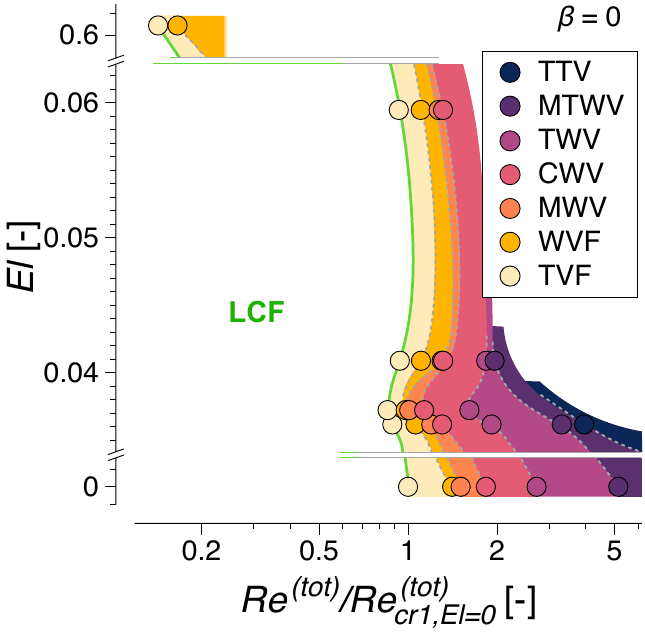}}
\subfigure[]{\includegraphics[width=0.49\textwidth]{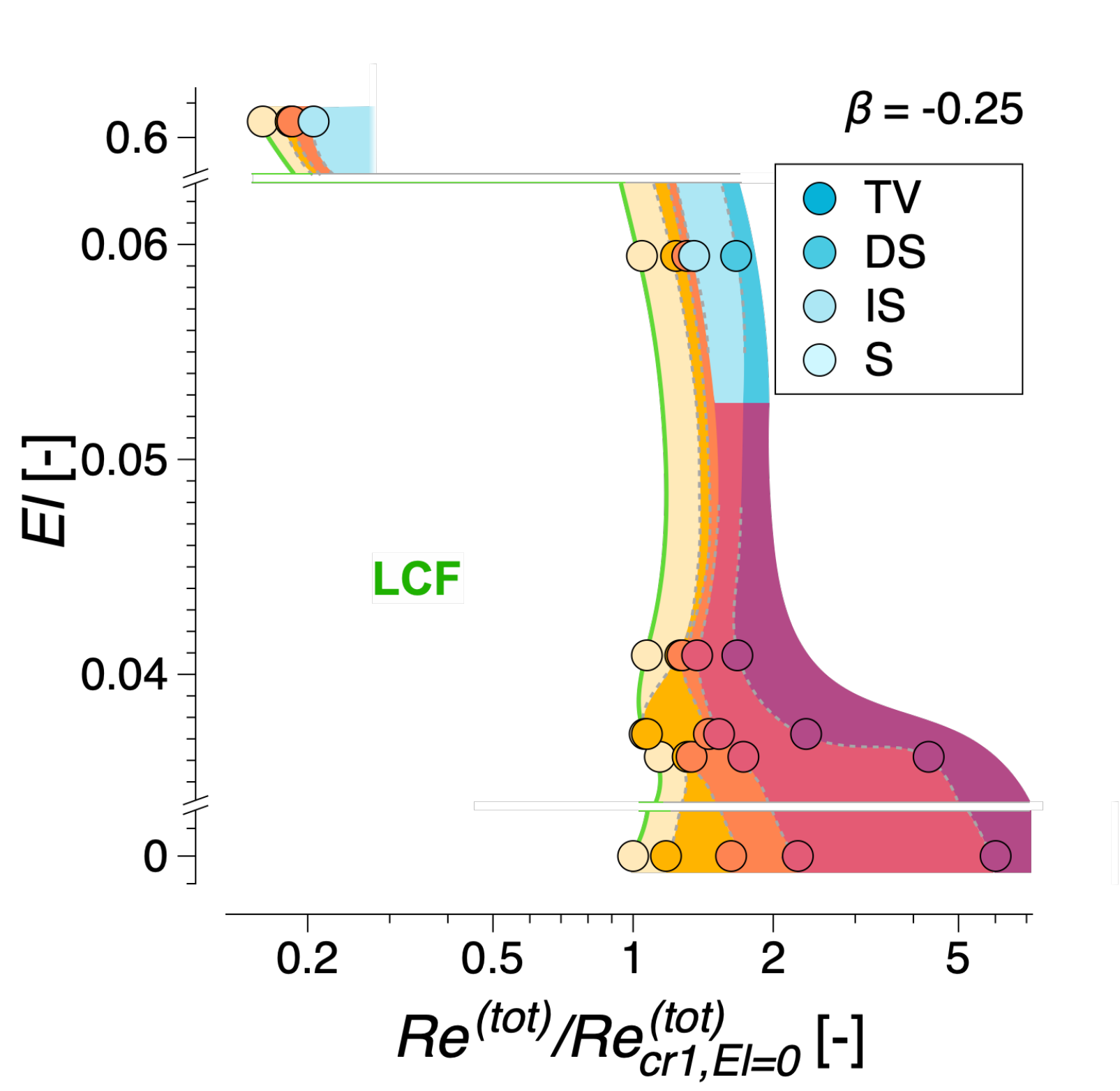}}
\subfigure[]{\includegraphics[width=0.49\textwidth]{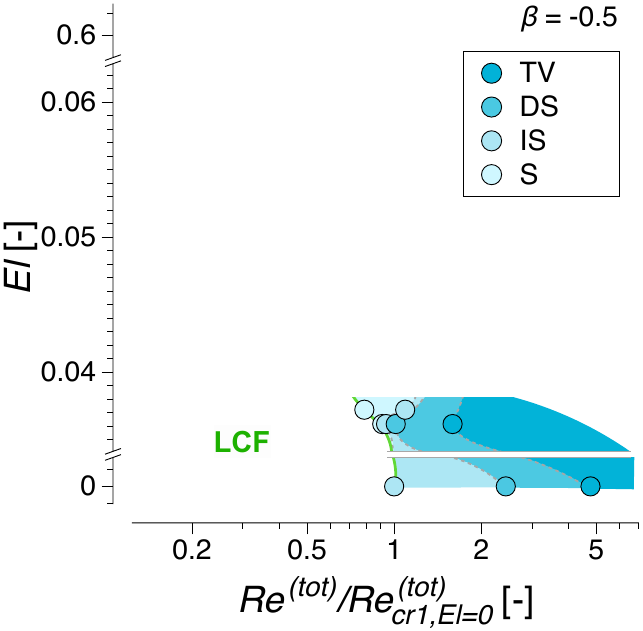}}
\caption{\label{fig12} Flow stability diagrams in the form of $El$ vs. $Re$ comparing the reference Newtonian case and the CNC suspensions for $\beta =$ (a) 0, (b) -0.25, (c) -0.5.}
\end{figure}

While Fig. \ref{fig10} displays the cascade and the variation of flow modes for relevant relative rotations of the cylinders and each CNC suspension, Fig. \ref{fig12} represents an overall view of the dynamics of flow transition in terms of the $El$ and $Re$ normalized by the critical $Re$ for the onset of the TVF regime for the Newtonian reference, i.e. $Re_{cr}/Re_{cr1, El=0}$, for $\beta=0$, -0.25 and -0.5 cases. We note that transition sequences for $\beta < -0.5$ have been omitted as they comprise cases where LCF is the only flow mode throughout the $Re, El$ investigated. Overall, for $El < 0.65$, in all cases, we can obtain the following relationship $Re_{cr1,El}/Re_{cr1,El=0} = 1 \pm 0.25$ (TVF). While initially causing a weak destabilization (the onset of instabilities occurs at lower $Re_{cr1}$ compared to the Newtonian case), see $\beta = 0, -0.5$ in Fig. \ref{fig12}(a) and (c), or a weak stabilization (the onset of instabilities occurs at higher $Re_{cr1}$ compared to the Newtonian case) of the flow, see $\beta = -0.25$ in Fig. \ref{fig12}(b), for $El = 0.65$ (3 wt\%), a pronounced destabilization of the flow is recorded with $Re_{cr}/Re_{cr1,El=0} < 0.2$ for $\beta = 0,-0.25$. Higher order transitions can experience more significant shifts and seem to be more prone to be destabilized by the addition of elasticity even for $\beta = -0.25$. This means that in the $\beta = 0$ case the TTV regime for example, while not captured in the Newtonian case, could be identified for $El = 0.036$. We note that for $\beta \neq 0$ the influence of elasticity does not equally affect supercritical instability modes, e.g. see for $El \in [0.036,0.059]$ TVF, WVF and MWV in Fig. \ref{fig12}(b) and S, IS in Fig. \ref{fig12}(c). In particular, the IS mode tends to destabilize as the CNC suspension reaches 1.0 wt\%. Later, the onset of IS is shifted to higher $Re/Re_{cr1, El=0}$. In contrast to IS, the spiral flow is absent in the Newtonian fluid, and it tends to emerge for the CNC with 1.0 and 1.5 wt\% with a shift to a lower critical $Re$. The DS and TV modes are present in water and the CNC with 1.0 wt\%, however, there are no visible traces of these modes in solutions with higher $El$.

%The dynamics of flow patterns at $\beta = -0.25$ reveal distinctions compared to $\beta = 0$. While the onset of CWV and TWV flow modes are shifted to lower $Re/Re_{cr1, El=0}$ values at increased $El$, the TVF and WVF transition experience stabilization for CNC 1.0 wt\% compared to water. However, the appearance of MWV demonstrates a minor deviation from the $\beta = 0$ with more inconsistency as the solution gets more concentrated. The spiral-featured modes only emerge at CNC with the highest $El$ with a stark destabilization shift from $El$ of 0.059 to 0.65, setting these distinguishable shifts apart from the rest of the flow transitions. This is plausibly due to the highly elastic nature of these solutions. 

\subsection{Elasticity and shear thinning in the Taylor-Couette flow of cellulose nanocrystal suspensions}

\begin{figure}[h!]
\subfigure[]{\includegraphics[width=0.49\textwidth]{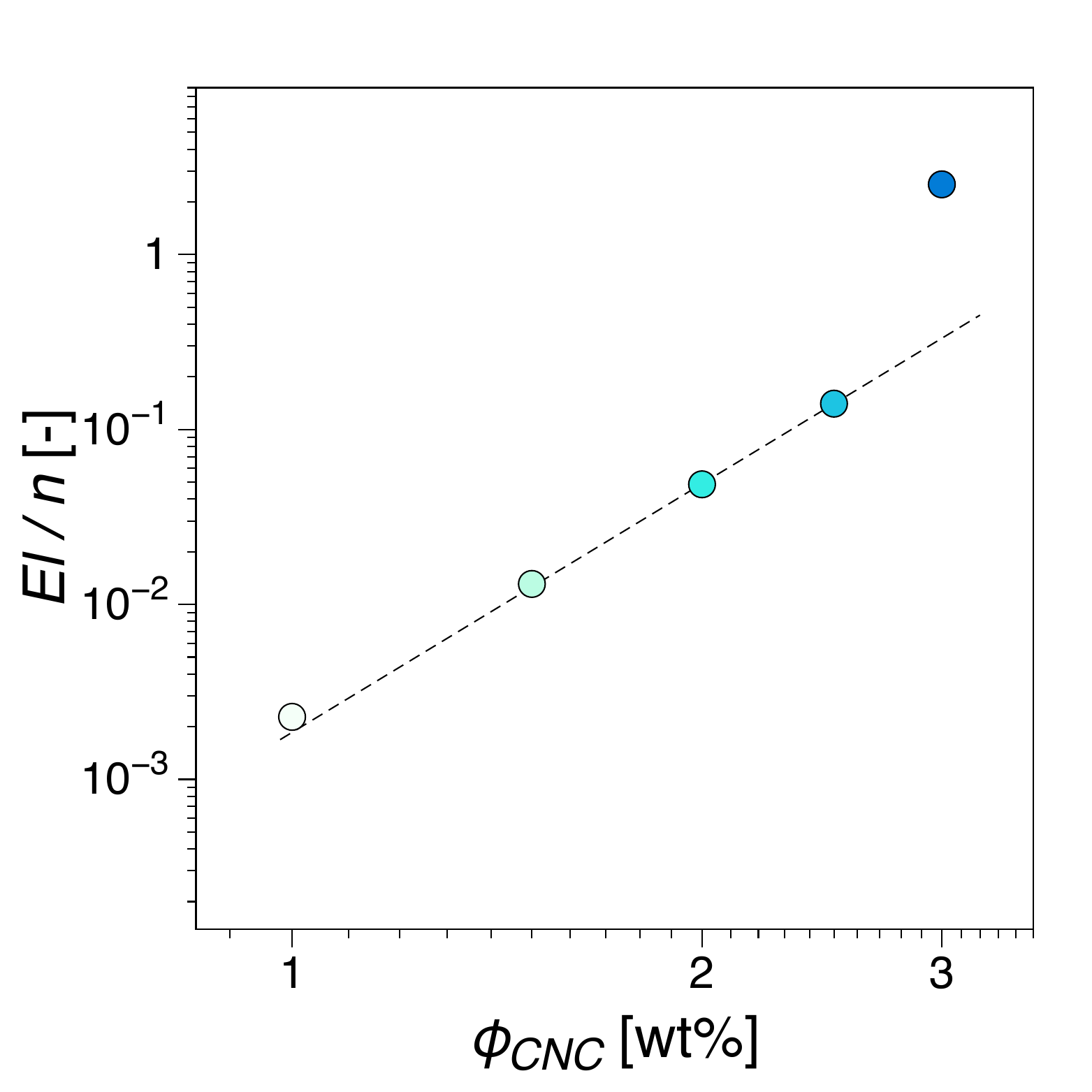}}
\subfigure[]{\includegraphics[width=0.49\textwidth]{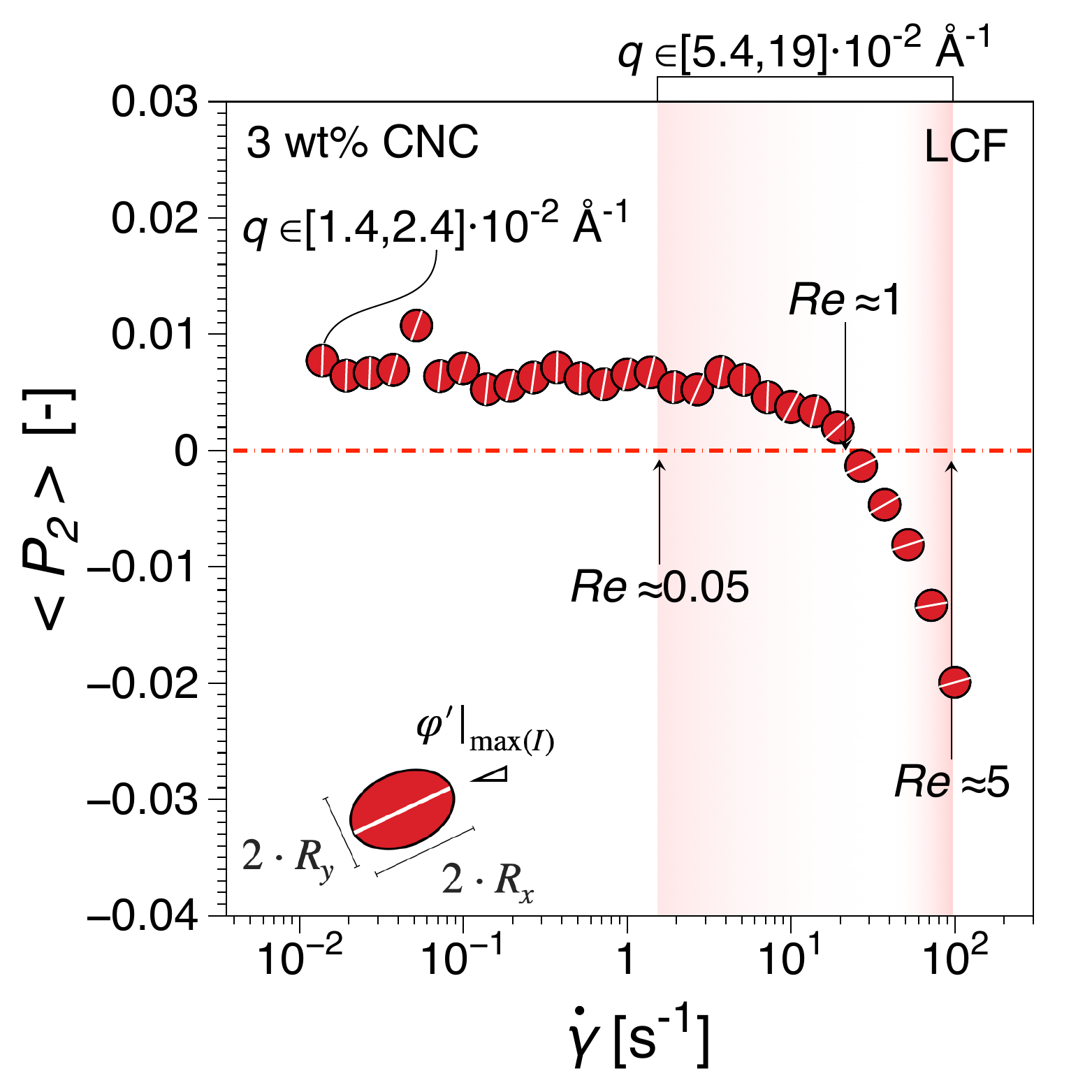}}
\caption{\label{figEln} (a) $El / n$ versus CNC concentration. (b) Hermans orientation parameter, $\left< P_2 \right>$, representation from rheo-SAXS experiments (LCF) as function of shear rate for the 3 wt\% CNC suspension. The datapoints plotted correspond to the azimuthal integration of the scattering intensity over $q \in [1.4,2.4]\cdot 10^{-2}$ Å$^{-1}$. The marked region corresponds to the shear rate / $Re$ range for orientation in the flow direction within $q \in [0.9,19]\cdot 10^{-2}$ Å$^{-1}$, estimation based on a broader range of experiments (not shown).}
\end{figure}

In this section we discuss the stability of CNC suspensions in the framework of elasticity and shear thinning effects. The Taylor-Couette flow of CNC suspensions presents a conundrum compared to other non-Newtonian cases in the scientific literature. While part of the discussion in the previous section has been carried out in the framework of $El=El(Re)$ stability diagrams, it is important to note that with increasing CNC concentration both elasticity, as quantified by $El$, and shear thinning, as quantified by $n_i$ show a power law dependence on the CNC concentration up to 2.5 wt\%, Fig. \ref{figEln}, see also Fig. \ref{parameters}. Thereafter the relative increase in $El$ exceeds considerably that of the $n_i$. However, the flow patterns remain Newtonian-like throughout the $El, n_i$ investigated. 

%The the spatial characteristics of the first instability mode, TVF, shows evidence of the presence of elasticity, and the critical parameters for the onset of instabilities are clearly shifted with increasing CNC concentration. However, no distinct non-Newtonian flow patterns could be identified and the destabilization / stabilization of the flow. Furthermore, the onset of instabilities is only weakly influenced by the increase in $El$ while higher order transitions are significantly affected.

The vast majority of non-Newtonian Taylor-Couette flow cases have been performed on polymer solutions. Such systems are inherently elastic due to the elastic nature of polymer chains. At the same time, polymeric solutions are also shear thinning, as the chains stretch and orient in the flow direction. To eliminate shear thinning, weakly elastic polymer solution compositions have been investigated, also known as Boger fluids, while more recently shear thinning fluids with different elasticity levels have been investigated. Thus, the general stability discussion on polymer solutions, from material point of view, revolves around the influences of elasticity and shear thinning. We note that the parameter space determining the TC flow stability includes not only material parameters but also setup parameters such as the relative counter-rotation of the cylinders, which is considered in the present study, as well as the radius and aspect ratios (0.91 and 34.4, respectively). Thus, despite a substantial number of studies on non-Newtonian TC flows available in the scientific literature, direct comparisons are not always possible. In terms of radius ratio, the present experiment can be considered as narrow-gap TC flow, similar to the works by Dutcher and Muller, but at almost half of their aspect ratio (60.7)\cite{Muller2013weak,Muller2013Moderate}. In contrast, the radius and aspect ratio used in the present study are considerably higher than some studies that are relevant for comparison in terms of elasticity and shear thinning (e.g. 0.776 and 21.56, respectively, by Balabani and co-workers\cite{Balabani2020TC, Balabani2021TC}. Thus, as we propose a different interpretation for the interplay between elasticity and shear thinning based on a mesophase (liquid crystalline domain) evolution with CNC concentration and $Re$, we will refer below mainly to flow and geometry cases that are similar to the present case. 

Dutcher and Muller have investigated the flow cascades of shear-thinning polymer solutions at low to intermediate $El$ at different relative cylinder rotations with flow phase diagrams similar to Fig. \ref{fig10} but in terms of critical Reynolds numbers \cite{Muller2013weak, Muller2013Moderate}. Our findings with $\beta = -0.25$ bear the closest similarity to their results at weak elasticities, that is, the CNC suspensions with $<2.5$ wt\%. However, our experiments show the onset of spiral-like flow modes in contrast to elastically inherent flow phases revealed in their study. In two independent contributions by Baumert and Muller, stationary counter-rotating vortices have been reported for non-shear-thinning fluid of dilute polyisobutylene solutions in polybutene of low elasticities ($El = 0.0562$\cite{MullerExotic1999} and 0.16 \cite{baumert1997flow}) for $\beta = -1$ \cite{MullerExotic1999}, where the former and latter case have similar elasticity level of the CNC suspension with CNC of 1.5-2.5 wt\%, and 3.0 wt\%, respectively. The same authors reported similar stationary counter-rotating vortices for highly elastic fluids of polymer solutions at $El= 1$ and $El= 44$, where depending on the viscosity, $Wi$, and $Re$, intermediate flow patterns such as Migrating Bands and Distorted counter-rotating vortices emerge \cite{baumert1997flow} for $\beta = -1$. 
Another work and based on linear stability analysis points to a flow mode called axisymmetric, oscillatory for the stationary inner cylinder at the elasticity level of 0.16 comparable to CNC of 3.0 wt\% \cite{AOM1992JOO}. However, in the current work and for stationary inner cylinder scenarios, LCF pattern persists to exist in the entire examined $Re$. Supercritical transitions from TVF to elastic rotating standing waves (RSW) which cascades to disordered oscillations (DO) \cite{groisman1996couette, groisman1998elastic} or the uninterrupted morphing of TVF into DO \cite{groisman1998elastic,steinberg1998elastic} at relatively higher $El$ values have also been reported.

The emergence of standing vortices (SV) \cite{Muller2013Moderate}, disordered rotating standing waves (DRSW) \cite{GroismanExotic1997}, elastically dominated turbulence (EDT) or elasto-inertial turbulence (EIT) \cite{Balabani2021TC,lacassagne2020vortex, Muller2013Moderate,gillissen2019_EIT} at moderate to high elasticity numbers for weakly shear-thinning polymer solutions have been ascribed to the significant contribution of the elastic response, with direct translation of LCF to EIT in certain cases \cite{Balabani2021TC, groisman1996couette}. In recent studies, it has been argued how the flow translations deviate from non-shear-thinning systems if the test fluid exhibits shear-thinning behavior \cite{Balabani2020TC, Balabani2021TC,cagney2019taylor,latrache2016defect,caton2006, bahrani_2015_shearthinning, topayev_2019_shearthinning,Crumeyrolle_2002}. In a work by N. Cagney et al, the authors have looked into the interplay of fluid's rheology and shear-thinning on flow instabilities \cite{Balabani2020TC}. Their experiments showed Newtonian-like flow patterns, however, with alteration in flow transition critical parameters. In another work, same authors have shown that both $Re_{cr1}$ and $Re_{cr2}$ ($\beta = 0$) tend to decrease non-monotonically with increasing $El$. $Re^{i}_{cr1}$ was relatively lower than the non-Newntonian fluids and a $Re^{i}_{cr2}$ significantly deviating from the Newtonian case \cite{cagney2019taylor}. Similar destabilizing effects of shear-thinning have also been discussed in other works both using experimental techniques \cite{Crumeyrolle_2002, Kim1997experimental,elcciccek2020effect,alibenyahia2012revisiting,elcciccek2020non, Jeff2018inertial} and numerical methods \cite{Ashrafi2011hearThinNum,coronado2004ShearThinNum,lockett1992ShearThinNum}. A few of studies have also argued that shear-thinning introduces asymmetric flow states such as Ribbons (RIB) and spiral vortex flow (SVF) \cite{caton2006, elcciccek2020non, Jeff2018inertial} absent in Newtonian fluids, or the expansion and contraction of vortices outwards and inwards in radial direction \cite{cagney2019taylor}, or modifications in waviness patterns attributed to the WVF states \cite{Balabani2020TC}.  In a study by Lacassagne et al, the interplay of shear-thinning and elasto-inertial contribution has been elaborated \cite{Balabani2021TC}, where the fluids with high shear-thinning levels and at moderate to high $El$ conventional Newtonian-like flow emerge with variations in wavenumber and the wave frequency of TVF and WVF \cite{Balabani2021TC}. The same work points to the emergence of elastically-regulated flow patterns only in low to moderate shear-thinning levels at high elasticities. %However, such elastically-dominated patterns disappear at high inertia irrespective of strongly elastic behaving fluid \cite{Balabani2021TC}. 

%We also need to distinguish the effect of elasticity and shear-thinning within liquid crystals systems like CNCs in this study and a typical polymer chain solution. In polymer solutions, normal stresses and subsequent elastically-dominated "hoop stresses" dictate the flow instability. In contrast, in CNC solutions at sufficiently high $Re$ beyond which chiral nematic or nematic assemblies are no longer present, such "hoop stresses" are absent, and the shear-thinning determines the flow transitions. 

Overall, these variations in flow modes reported for polymer solution highlight the distinctive nature of CNC suspensions TC flow stability. For the weakly elastic CNC suspensions, $\phi_{CNC} \in [1,2.5]$ ($El<<1$), it could be argued that flow stability is either dominated by shear-thinning, or the Newtonian infinite shear viscosity plateau has been reached. Although the latter is not identifiable in the viscosity functions this can be inferred based on the shear thinning slope and the fact that $\eta_\infty$ should be greater than $10^{-3}$ (water). Within the weakly elastic range, with increasing CNC concentration, both $El$ and $n_i$ gradually increase, see Fig. \ref{figEln}(a), meaning that the conditions for the development of elastically-driven novel instability modes are not met. However, the fact that even for 1 wt\% CNC $\kappa$ is significantly lower than the Newtonian case for all $Re$ could suggest that shear thinning does influence flow stability up to the highest $Re$ investigated. This would mean that the transition sequence recorded for $\geq 2$ wt\% CNC from TVF $\rightarrow$ WVF $\rightarrow$ MWV $\rightarrow$ IS is also a consequence of shear thinning.

For the moderate $El$ case, $\phi_{CNC} = 3$ wt\%, there is a significant increase in both $El$ and $n_i$. In this case, again, the conditions for elastically-induce patterns are most likely not met, or elastic effects are suppressed by the flow. The fact that there are no clear fingerprint patterns that are associated to the chiral nematic phase present in the POMs suggests that the systems contain a level of agglomeration (here by agglomerates we mean clusters of CNCs without positional and/or orientational order at lengthscales greater than the particle lengthscale). To gain a more profound insight into the structural dynamics of the flow, we have performed rheo-SAXS experiments on the 3 wt\% CNC sample, Fig. \ref{figEln}(b). At low shear rates we see evidence of vertical orientation, $\left< P_2 \right> \; > 0$. Similar results have been shown by Pignon et al. \cite{Pignon2021}; this has been broadly associated to the orientation of tactoids containing liquid crystalline domains with preferential orientation perpendicular to the flow direction. The loss of vertical orientation based on the shear rate in the rheo-SAXS test would correspond to $Re \approx 1$ in the TC flow case. This has been associated to the breakup of the meshophase into nematic and then individual CNCs that orient in the flow direction\cite{Pignon2021}. We note, that depending on the preparation method the shear rate corresponding to the orientation in the flow direction can vary considerably depending on the $q$-range investigated. Thus, based on several similar experiments over a broader $q$-range (data not shown) we estimate that for 3 wt\% a reasonable $Re$-range would be between 0.5 and 5. This is considerably lower than $Re_{cr1}$ for 3 wt\% CNC, $\beta = 0$, see Fig. \ref{figAll2}. This means that the pre-TVF instability observed can be still a form of shear thinning effect. Alternatively, due to the agglomerates apparent in the 3 wt\% it is possible that some parts of the mesophase can most likely form a mesophase under the influence of shear. At this point without further evidence this remains a speculation. We also need to distinguish the effect of elasticity and shear-thinning within liquid crystals systems like CNCs in this study and a typical polymer chain solution. In polymer solutions, normal stresses and subsequent elastically-dominated "hoop stresses" dictate the flow instability \cite{ghanbari_2014}. In contrast, in CNC solutions at sufficiently high $Re$ beyond which chiral nematic or nematic assemblies are no longer present, such "hoop stresses" are absent, and the shear-thinning determines the flow transitions. 

%It is also important to distinguish the effect of elasticity and shear-thinning within liquid crystals systems like CNCs in this study and a typical polymer chain solution.  WhiIn polymer solutions, normal stresses and subsequent elastically-dominated "hoop stresses" dictate the flow instability. In contrast, in CNC solutions and as $Re$ is elevated, such "hoop stresses" are absent due to the disruption of the liquid crystalline mesophase and thus shear-thinning continues to determine the flow transitions. 

Finally, we note that the flow stability of CNC suspensions could be further elucidated without the requirement for higher CNC concentrations by employing different preparation methods, which can significantly influence their self-assembly and rheological properties\cite{Kadar2023}.

\section{Summary and conclusions}

In the current work, the flow stability of CNC suspensions at different concentrations corresponding to altered elasticities has been investigated. Specifically, a novel rheo-optical method has been utilized to directly observe flow patterns at altered rotations of inner and outer cylinders of the TC geometry. The particular liquid crystalline nature of the cellulose nanocrystal and how they respond to polarized light can explain the applicability of such a method. The supercritical flow modes at varied relative cylinder rotations have been observed and detailed in light of the material functions and elastic response of the cellulose nanorods. In essence, the flow transitions observed follow a Newtonian-like sequence but modified mainly by shear-thinning. In brief, the Taylor-Couette flow stability problem for CNC solutions is significantly influenced by the strong shear-thinning behavior of such solutions in non-linear flow conditions, where the elasticity dwindles due to the disengagement of nematic or chiral nematic domains. Also, our findings on the one hand sheds light on distinctions between liquid crystalline and polymer chains in terms of their viscoelastic nature, and on the other hand, evidences the understanding the concept of elasticity beyond a single number used for defining the viscoelastic response of a Non-Newtonian fluid.

\begin{acknowledgments}
The work was supported by Chalmers Foundation though the project Chalmers Center for Advanced Neutron and X-ray scattering techniques. The “FibRe - Competence Centre for Design for Circularity: Lignocellulose- based Thermoplastics” partly funded by the Swedish Innovation Agency VINNOVA (Grant Number 2019-00047), Chalmers Area of Advance Materials Science and the Wallenberg Wood Science Centre (WWSC) are also gratefully acknowledged. The authors are grateful to Anton Paar GmbH for supporting the development of the custom Taylor-Couette geometries. We acknowledge MAX IV Laboratory for time on Beamline CoSAXS under Proposals 20200458 and 20221436. Research conducted at MAX IV, a Swedish national user facility, is supported by the Swedish Research council under contract 2018-07152, the Swedish Governmental Agency for Innovation Systems under contract 2018-04969, and Formas under contract 2019-02496.
\end{acknowledgments}

\section*{Data Availability Statement}
All data is available on request from the authors.

\bibliography{TC}% Produces the bibliography via BibTeX.

\clearpage

\appendix

%\section{Additional information on experiment and procedures}

\section*{Supplementary information}\label{A}

%\begin{figure}[h!]
%\centering
%\includegraphics[width=0.325\textwidth]{_Fig3.jpg}
%\caption{Schematic illustration of a biphasic CNC suspension: (i) nematic and/or (ii) chiral nematic orders that comprise the liquid crystalline phase; the rest of the domain (green) comprise the so-called isotropic phase.} \label{fig3}
%\end{figure}

\begin{figure}[h]
\includegraphics[width=0.48\textwidth]{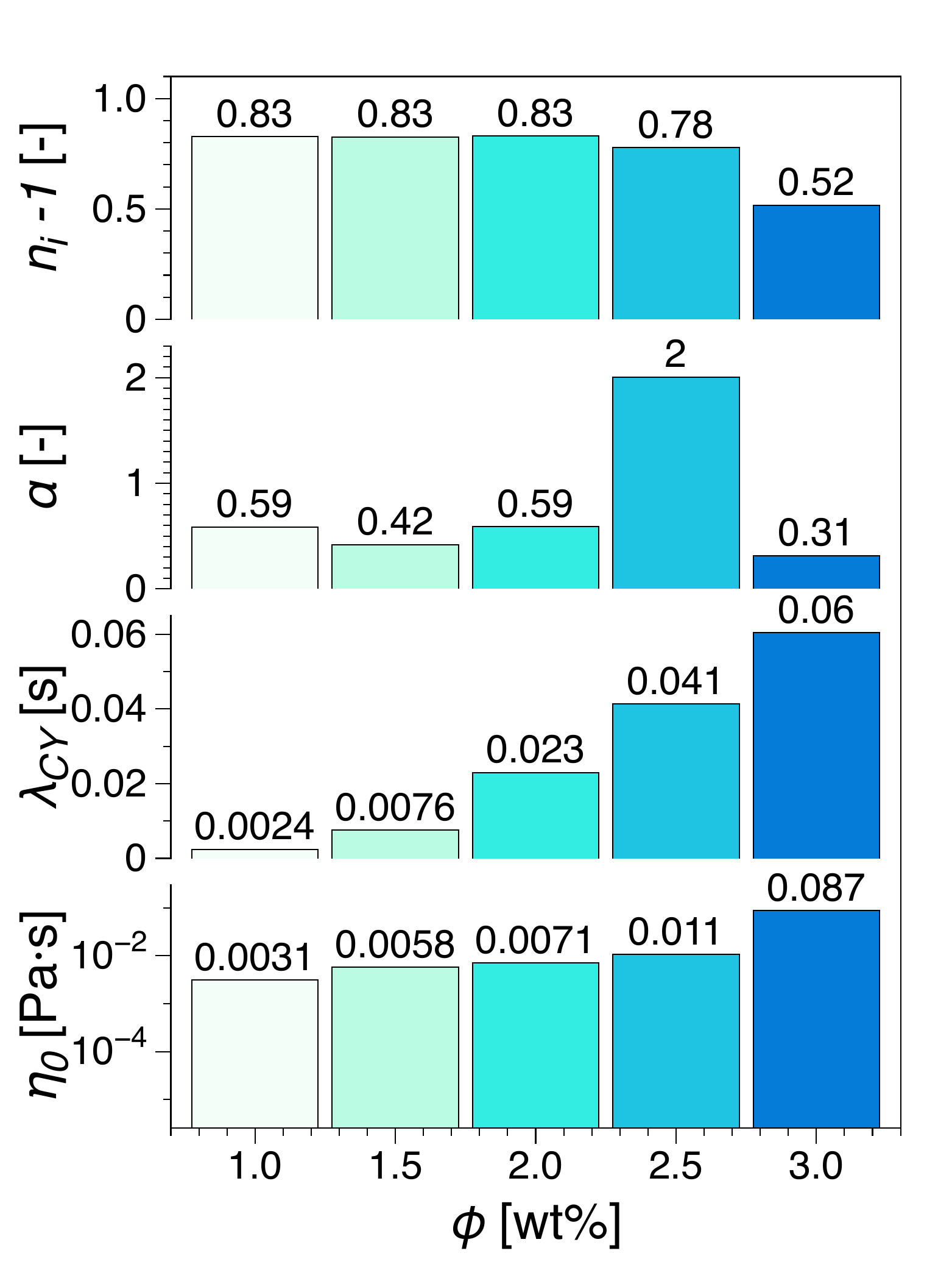}
\caption{\label{fits_CY} Carreau-Yasuda, Eq. (1), fit parameters.}
\end{figure}

\begin{figure}[t]
\subfigure[]{\includegraphics[width=0.47\textwidth]{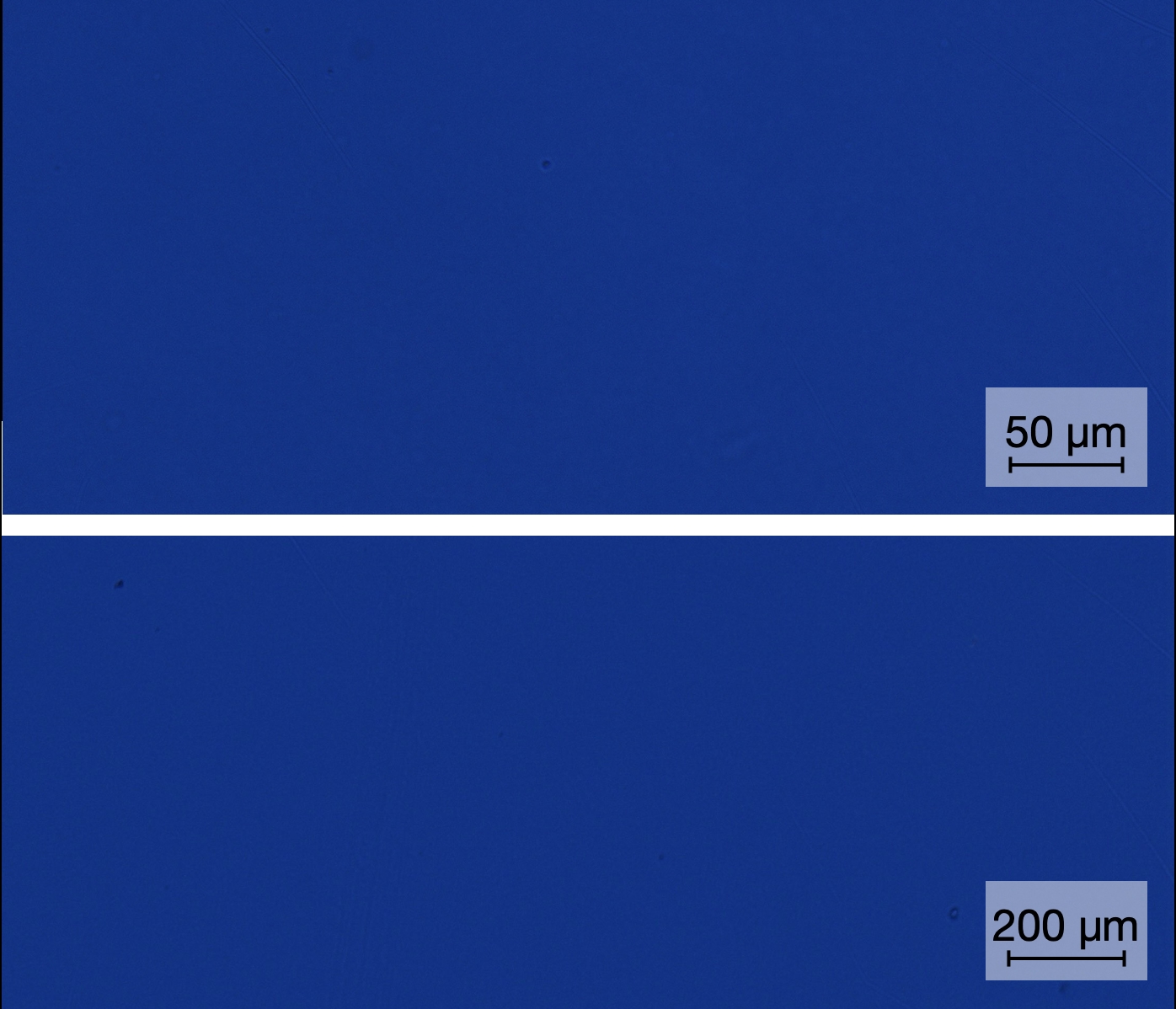}}
\qquad
\subfigure[]{\includegraphics[width=0.47\textwidth]{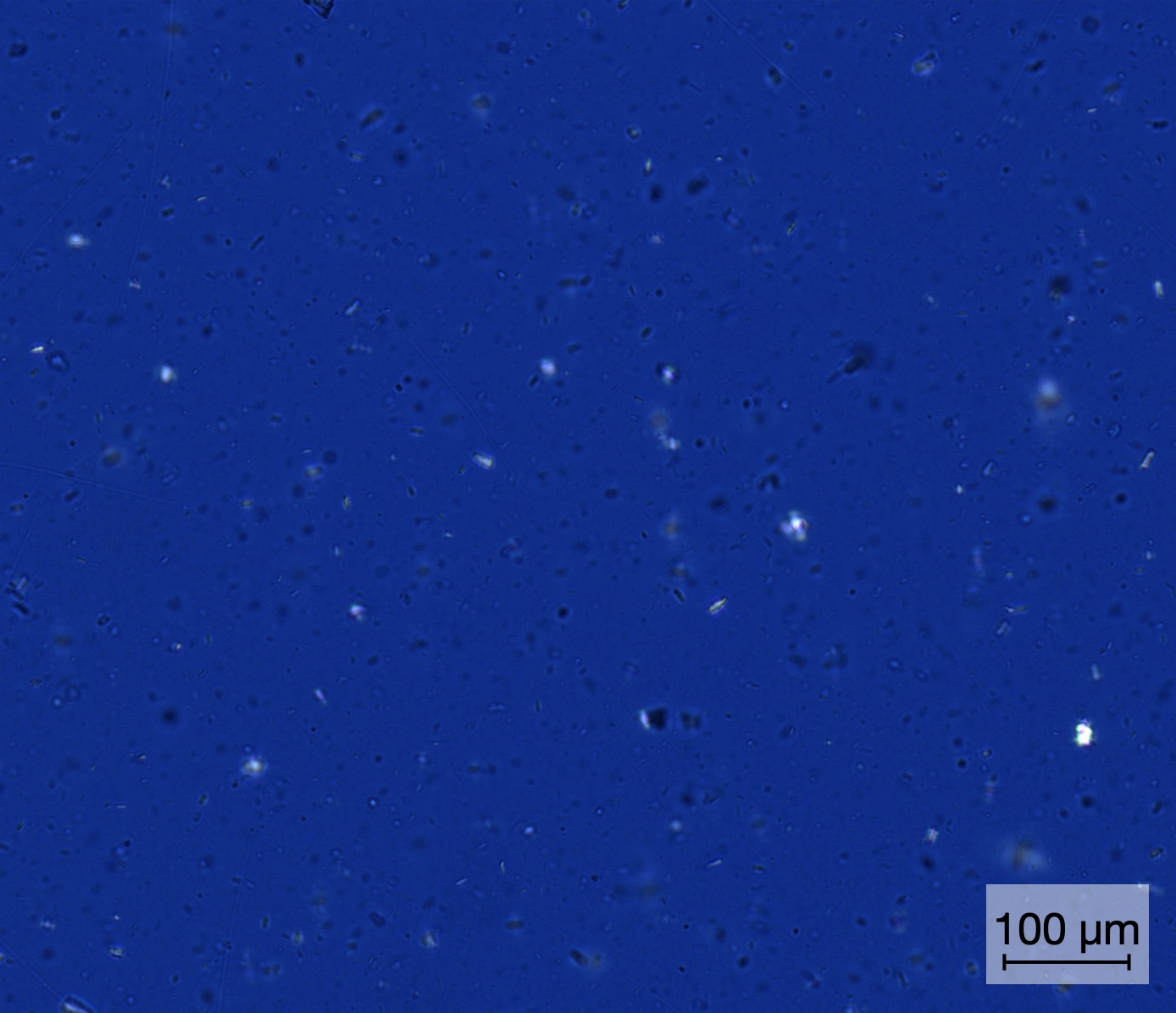}}
\caption{\label{POM} Polarized optical microscopy (cross-polarized) images of (a) 2wt\% and (b) 3wt\% CNC suspensions.}
\end{figure}

\begin{figure}[t]
\includegraphics[width=\textwidth]{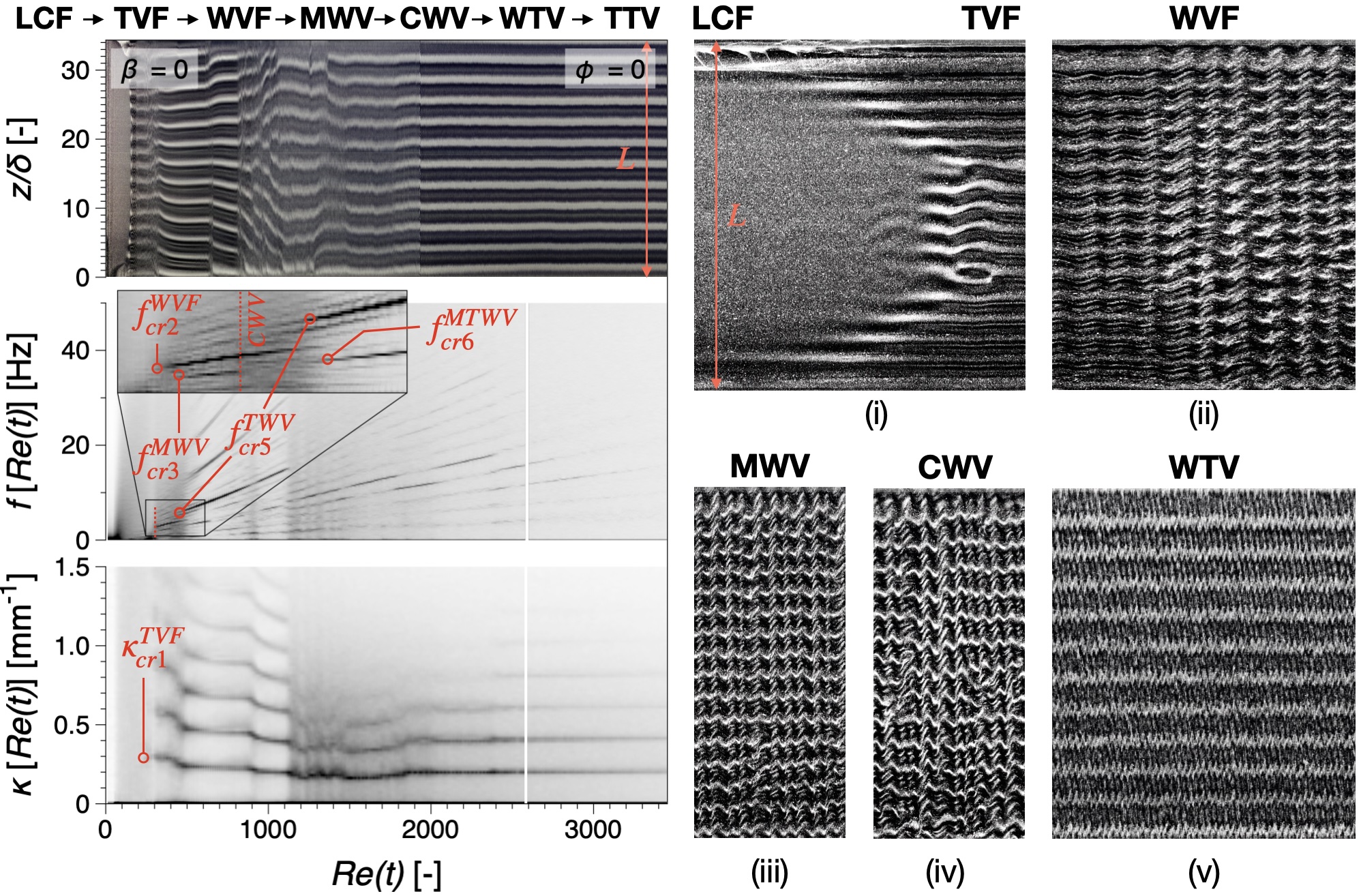}
\caption{\label{figB1a} Transition sequences for the Newtonian reference sample ($\phi = 0$ wt\% CNC; water with visualization particles): (a) $\beta = 0$, (b) $\beta = -0.5$ and (c) $\beta = -1$.}
\end{figure}

%\begin{figure}[p]
%\subfigure[]{\includegraphics[width=0.48\textwidth]{_Fig_Torque_a.pdf}}
%\subfigure[]{\includegraphics[width=0.48\textwidth]{_Fig_Torque_b.pdf}}
%\subfigure[]{\includegraphics[width=0.48\textwidth]{_Fig_Torque_c.pdf}}
%\subfigure[]{\includegraphics[width=0.48\textwidth]{_Fig_Torque_d.pdf}}
%\subfigure[]{\includegraphics[width=0.48\textwidth]{_Fig_Torque_e.pdf}}
%\caption{\label{fig8} Torque evolution with $Re_{tot}$ for (a) $\beta=0$, (b) $\beta=-0.25$, (c) $\beta=-0.5$, (d) $\beta=-0.75$, (e) $\beta=-1$ }
%\end{figure}

\begin{figure}[p]
\subfigure{(a)}{\includegraphics[height=6.3cm]{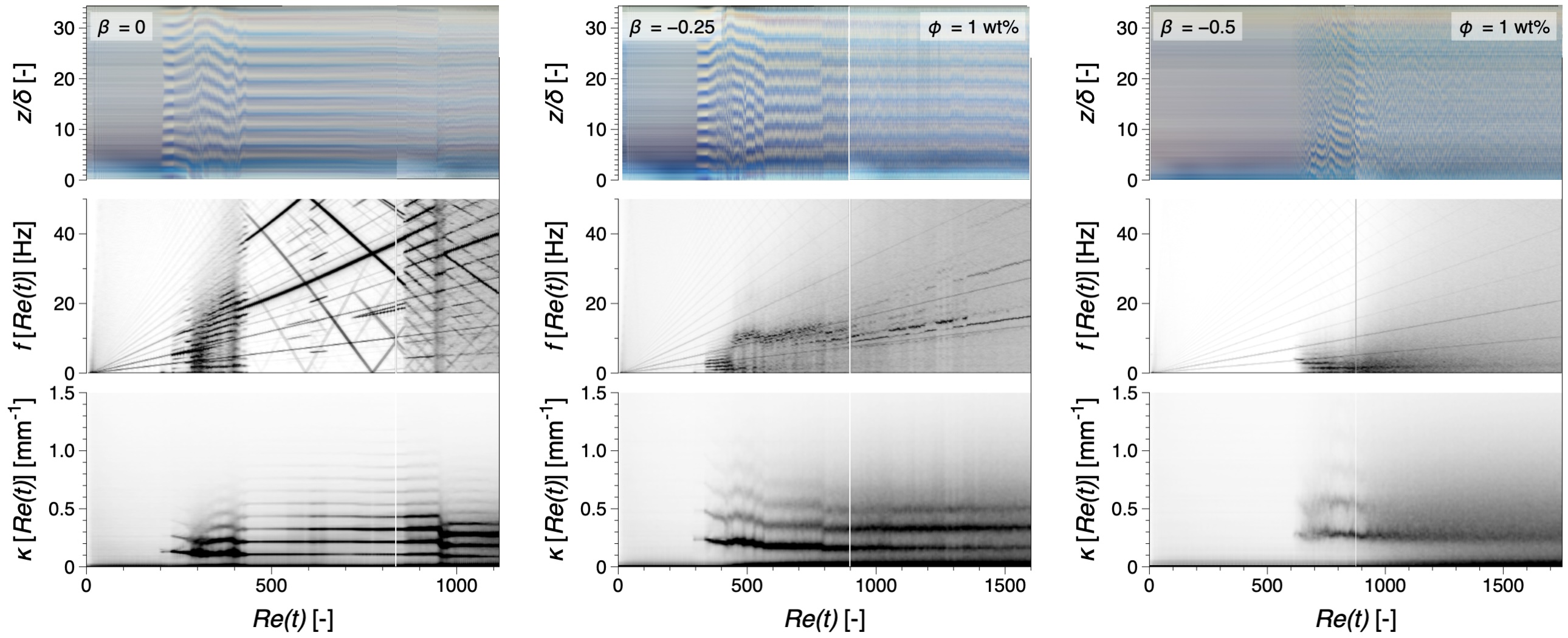}}
\subfigure{(b)}{\includegraphics[height=6.3cm]{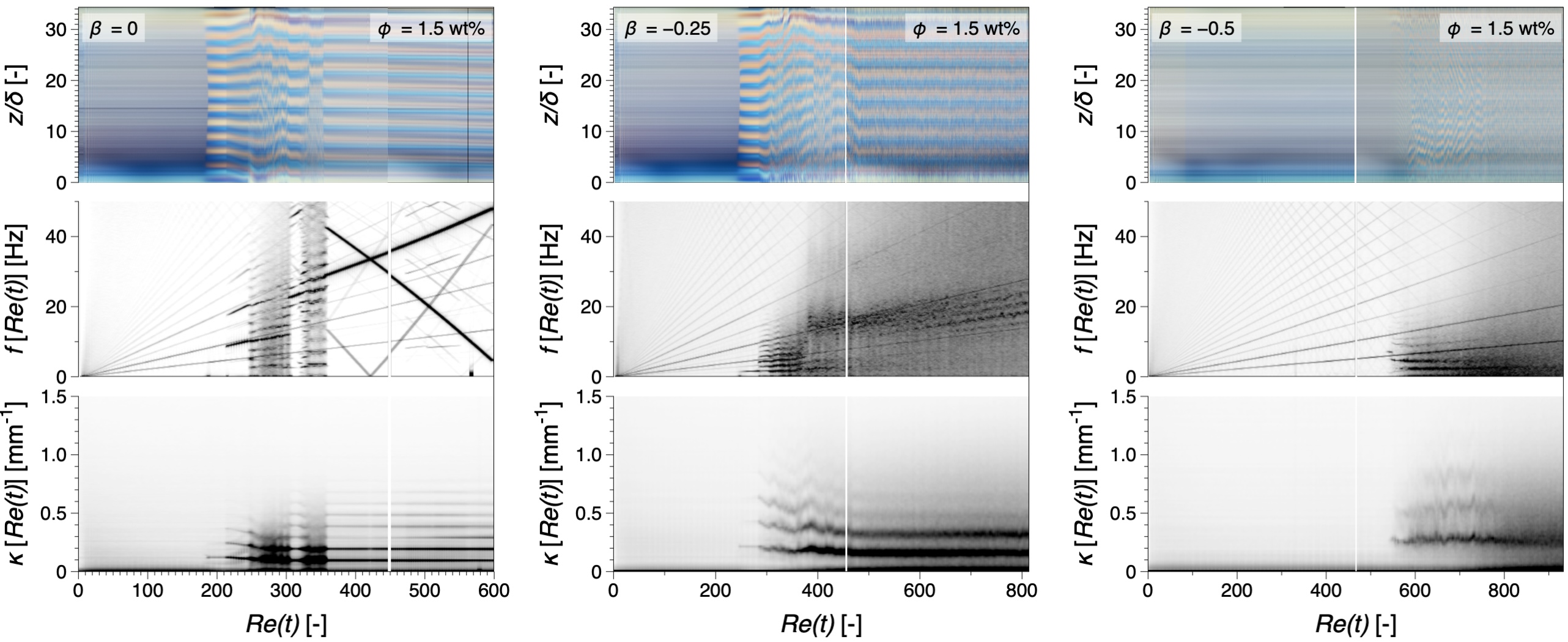}}
\subfigure{(c)}{\includegraphics[height=6.3cm]{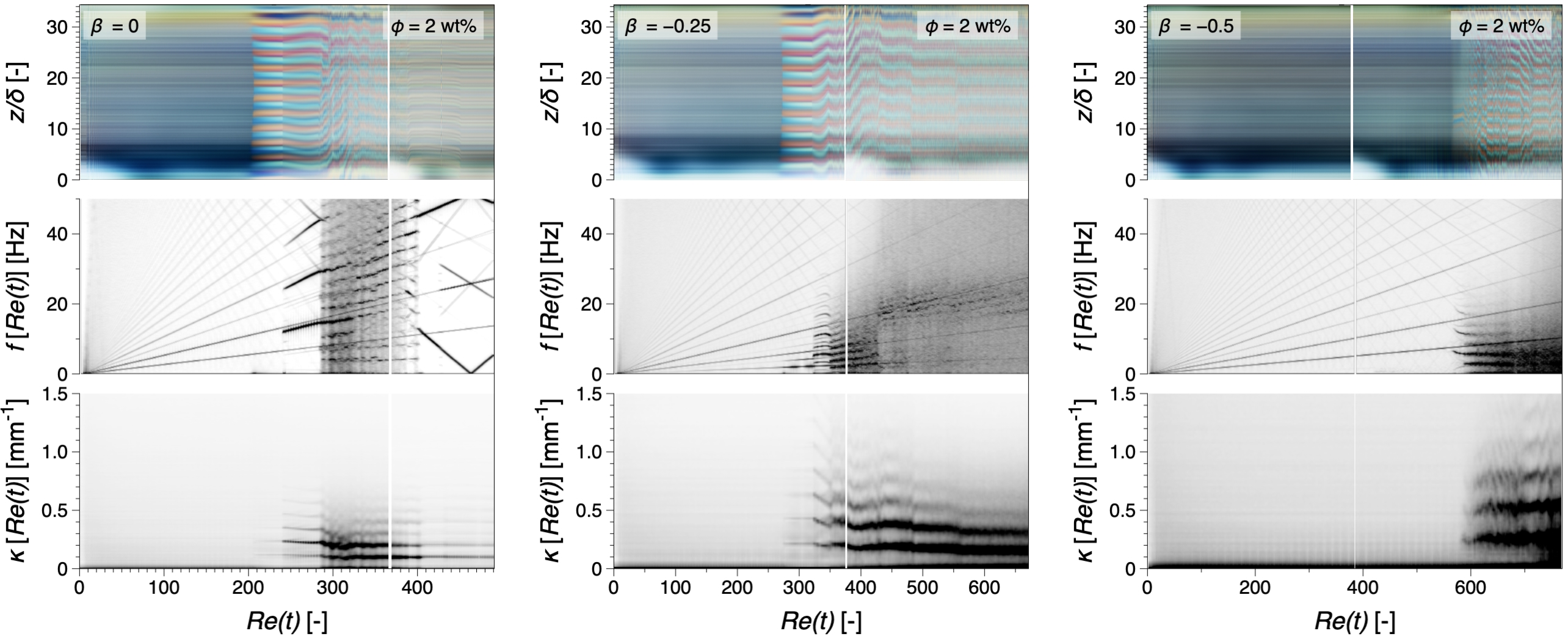}}
\text{(Continued on next page)}
\end{figure}

\begin{figure}[p!]
\subfigure{(d)}{\includegraphics[height=6.3cm]{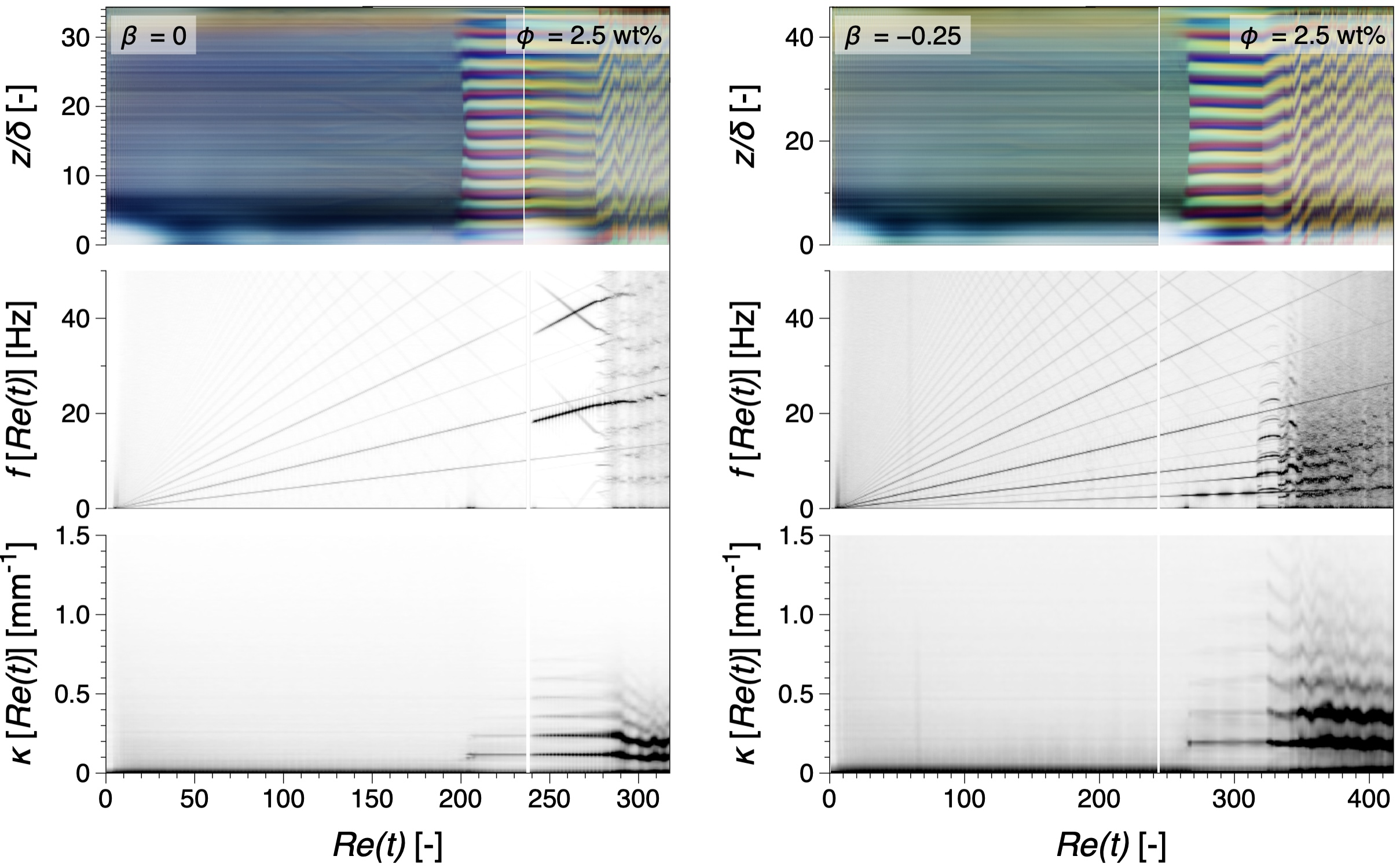}}
\subfigure{(e)}{\includegraphics[height=6.3cm]{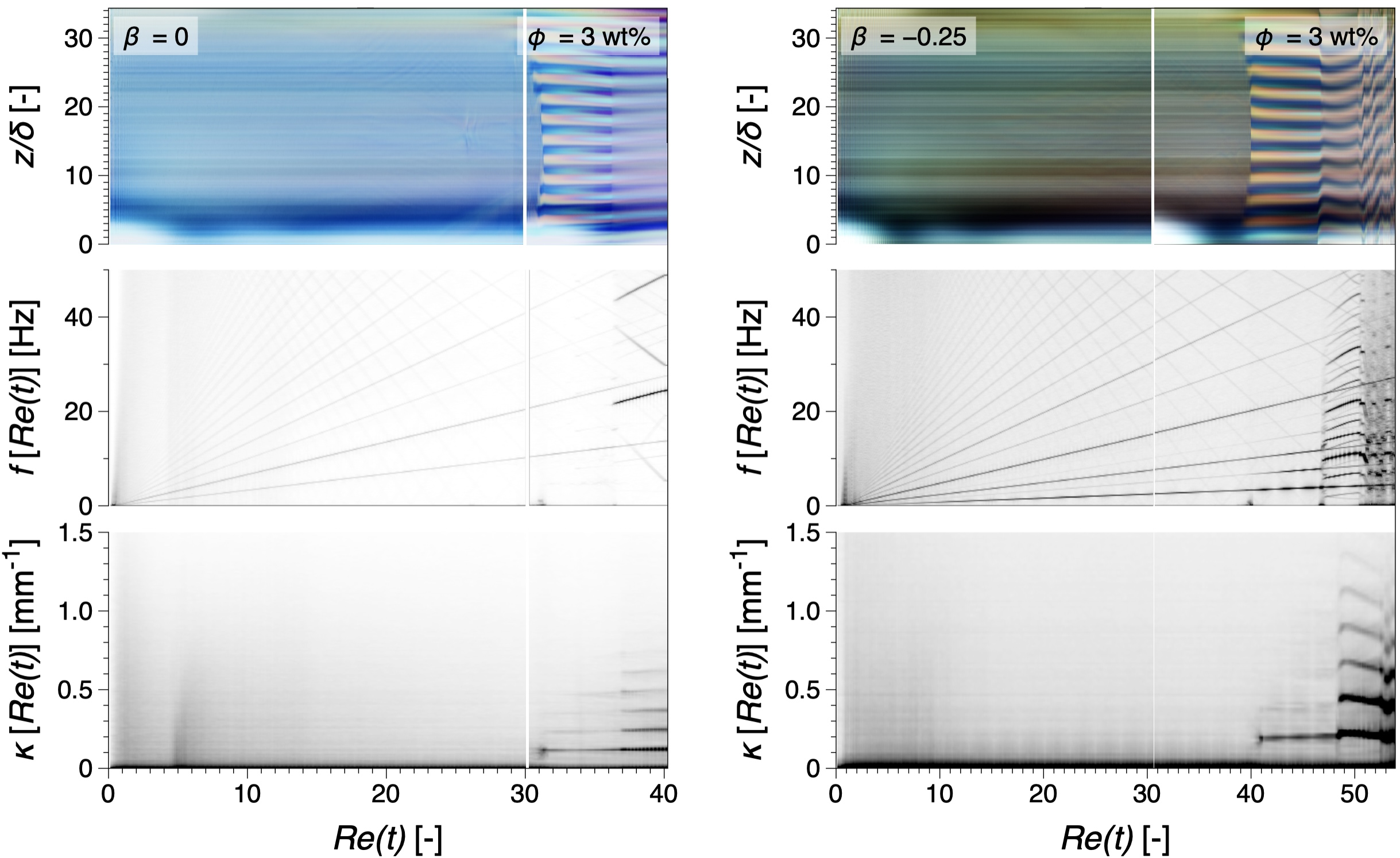}}
\caption{\label{figAll2} Compilation of space-time diagrams and spectral dynamics for all CNC concentrations and all relative cylinder rotation cases that showed instabilities.}
\end{figure}

\clearpage

\subsection[S1]{Hermans orientation parameter representation}

Each point in the Hermans orientation parameter, $\left< P_2 \right>$, is scaled according with magnitude and orientation of the microstructure. Within the integration limits used, the resulting orientation factor is $\left< P_2 \right> \in [1,-0.5]$. Therefore, $R_x$ and $R_y$ in Fig. \ref{figEln}(b), are defined as:
\begin{equation}
    R_x = 1-|\left< P_2 \right>|\label{Rx}
\end{equation}
\begin{equation}
    R_y = 2(0.5-|\left< P_2 \right>|) \label{Ry}
\end{equation}
The ellipses thus constructed are rotated with the angle $\phi'$ corresponding to the orientation of the nanoparticle in real-space.

\end{document}